\documentclass[fleqn,usenatbib]{mnras}
\pdfoutput=1

\usepackage{newtxtext,newtxmath}

\usepackage[T1]{fontenc}
\usepackage{ae,aecompl}

\usepackage{graphicx}	\usepackage{amsmath}	\usepackage{amssymb}	\usepackage{longtable} \usepackage{color, colortbl} \usepackage{lscape} \usepackage{rotating} 
\definecolor{Gray}{gray}{0.9}
\definecolor{white}{rgb}{1.0, 1.0, 1.0}

\title[AGN in Dwarf Galaxies]{X-ray detected AGN in SDSS Dwarf Galaxies}

\author[K. L. Birchall et al.]{
Keir L. Birchall,\thanks{E-mail: klb64@leicester.ac.uk}
M. G. Watson,
and J. Aird
\\
Department of Physics \& Astronomy, University of Leicester, University Road, Leicester LE1 7RH, UK
}

\date{Accepted XXX. Received YYY; in original form ZZZ}

\pubyear{2020}

\begin{document}
\label{firstpage}
\pagerange{\pageref{firstpage}--\pageref{lastpage}}
\maketitle

\begin{abstract}
In this work we present a robust quantification of X-ray selected AGN in local ($z \leq 0.25$) dwarf galaxies ($M_\mathrm{*} \leq 3 \times 10^9 \mathrm{M_\odot}$). We define a parent sample of 4,331 dwarf galaxies found within the footprint of both the MPA-JHU galaxy catalogue (based on SDSS DR8) and 3XMM DR7, performed a careful review of the data to remove misidentifications and produced a sample of 61 dwarf galaxies that exhibit nuclear X-ray activity indicative of an AGN. 
We examine the optical emission line ratios of our X-ray selected sample and find that optical AGN diagnostics fail to identify 85\% of the sources. We then calculated the growth rates of the black holes powering our AGN in terms of their specific accretion rates ($\propto L_\mathrm{X}/M_\mathrm{*}$, an approximate tracer of the Eddington ratio). Within our observed sample, we found a wide range of specific accretion rates. 
After correcting the observed sample for the varying sensitivity of 3XMM, we found further evidence for a wide range of X-ray luminosities and specific accretion rates, described by a power law. Using this corrected AGN sample we also define an AGN fraction describing their relative incidence within the parent sample. We found the AGN fraction increases with host galaxy mass (up to $\approx$ 6\%) for galaxies with X-ray luminosities between $10^{39} \mathrm{erg/s}$ and $10^{42} \mathrm{erg/s}$, and by extrapolating the power law to higher luminosities, we found evidence to suggest the fraction of luminous AGN ($L_\mathrm{X} \geq 10^{42.4} \mathrm{erg/s}$) is constant out to $z \approx 0.7$.
\end{abstract}


\section{Introduction}
\label{sec:intro}
Supermassive black holes (SMBHs) have been known to exist at the centre of massive galaxies with a bulge for some time \citep{KormendyHo13} however as one goes down the mass scale, the ubiquity of such objects becomes more uncertain. \\

Quantifying the fraction of dwarf galaxies with an active galactic nucleus (AGN) allows a lower limit to be established on the fraction that contain black holes. Some dwarf galaxies have much lower metallicity relative to their high mass counterparts and have not tidally interacted with their neighbours meaning they can be considered an analogue for galaxies in the high redshift Universe \citep{Bellovary11}. Thus we can use information about their AGN content to provide insights into the possible mechanisms that form the seeds of supermassive black holes in the the very early universe. 
Black hole formation from stellar collapse is possible at all redshifts but the early universe contained low metallicity gas in abundance which facilitated the growth of more massive Population III stars. After collapsing into stellar mass black hole seeds \citep{Madau01} they could have undergone a series of intermittent super-Eddington accretion episodes \citep{Madau14} or merged with other seeds to form more massive black holes \citep{MillerDavies12}. Inefficient gas cooling can also prevent star formation from occurring, thus allowing black holes to form from the the direct collapse of a proto-galactic dust cloud \citep{Begelman06}. Crucially, however the low angular momentum gas required for this method of formation only existed in sufficient quantities in the early Universe. As disk-like structures began to form, the gas within would gain angular momentum and become less likely to collapse and form a black hole. Thus if direct collapse were the dominant seeding mechanism then the fraction of dwarf galaxies hosting black holes is expected to be much lower than if the Population III stellar seeding mechanism was dominant \citep{Greene2012}. For a more in-depth review of black hole formation mechanisms see the review by \cite{LatifFerrara16}.\\

Black holes can be identified within massive galaxies using dynamical methods, however current technological limits mean these methods are difficult to apply at lower galactic masses. Consequently, more indirect approaches have to be adopted. Identifying the presence of AGN in dwarf galaxies has become an area of growing interest to assess the black hole population in dwarf galaxies. There is increasing evidence, across a range of wavelengths to show that at least a small number of AGN do exist within this mass range (see review by \cite{ReinesComastri16}). Some of the first large-scale studies into this area focused on optical emission \citep{ReinesGreenGeha13} identifying black holes through a combination of identifying broad line emission and measuring line strength. However, the effectiveness of this method can be limited by obscuration from dust or star formation signatures. 

AGN emission in the X-ray band generally dominates over other sources and is less easily obscured. A number of studies have had success in using this emission to identify the presence of AGN in dwarf galaxies (e.g. \cite{Reines11, Lemons15, Pardo16, Paggi16, Mezuca16, Mezcua18}). However, the techniques employed are not free from problems. Lower mass galaxies, like the ones being studied, tend to host lower mass central black holes hence less luminous emission is expected. This difficulty can also be compounded if the AGN being studied are very weakly accreting. The biggest challenge for studies like these is to rule out emission from other potential X-ray sources.\\

This paper presents one of the first large-scale and robust quantification of AGN in local dwarf galaxies. Our work uses an unbiased sample, being based on the XMM-Newton Serendipitous X-ray Survey and the SDSS which covers a large sky area. First we describe the construction of such a sample of local ($z \leq 0.25$) dwarf galaxies that possibly host AGN (section \ref{sec:data}). Next we analyse the properties of that emission to refine our sample to the most likely hosts (section \ref{sec:emissionAnalysis}). To better understand the AGN identified this way and the environments they inhabit we perform a series of follow-up measurements. We compare the effectiveness of optical identification methods to see if they correctly identify our X-ray selected AGN hosts (section \ref{sec:BPT}). Next, to gain more of an insight into the environment of our AGN, we then calculate the accretion rates of their central SMBHs (section \ref{sec:sBHAR}). Finally, we characterise the variable sensitivity limits of 3XMM, from which we can better understand the underlying distribution of AGN in dwarf galaxies (section \ref{sec:UL}). This allows us to investigate the population's AGN fraction as a function of host galaxy mass and redshift (section \ref{sec:fraction}). Throughout, we assume Friedman-Robertson-Walker cosmology in this paper with $\Omega = 0.3$, $\Lambda = 0.7$ and $H_0 = 70\mathrm{\ km\ s^{- 1} Mpc}^{- 1}$. 

\section{Constructing the Dwarf Galaxy Sample}
\label{sec:data}

What follows is a description of the construction of a sample of dwarf galaxies potentially hosting an AGN. First, a parent galaxy sample is isolated from the MPA-JHU catalogue, an SDSS value-added catalogue\footnote{Available at \href{http://www.mpa-garching.mpg.de/SDSS/DR7/}{http://www.mpa-garching.mpg.de/SDSS/DR7/}}. Next, these objects are position matched to serendipitous sources found in 3XMM \citep{Rosen16} and followed up with a visual assessment.

\subsection{MPA-JHU}
Optical photometry and spectroscopy covering 9274 $\mathrm{deg^2}$ of the sky can be found in the Sloan Digital Sky Survey Data Release 8 (SDSS DR8). The MPA-JHU catalogue provides derived estimates of galaxy properties such as stellar mass, star formation rate (SFR) and emission line fluxes, for 1,472,583 objects in this release. 

Stellar masses are provided by the MPA-JHU catalogue whose method is based on that described in \cite{Kauffmann03}.The MPA-JHU catalogue analysis uses template spectra made from a linear combination of single stellar population models generated using the \cite{BruzualCharlot03} code. These models consider 10 possible ages - from 0.005 to 10 Gyr - and 4 possible metallicites - from 0.25 to 2.4 $\mathrm{Z_\odot}$. They model galaxies  as a single metallicity population with the chosen model being the one that yields the minimum $\chi^2$. They then subtract this from the observed spectrum and the remaining emission lines are modelled as Gaussians. 
Rather than simply taking the model with the best $\chi^2$, it employs a Monte-Carlo fitting technique and produces a probability distribution for each observed and subsequently calculated property; the most likely value being the median of the distribution. 
Given the size of the SDSS fibre aperture, the spectral flux measurements required by \cite{Kauffmann03} would be dominated by light from the galactic centre if based on SDSS spectroscopy. Instead, the ugriz photometry from the full extent of the galaxy is used, and the total stellar mass is calculated by fitting to model magnitudes. A \cite{Kroupa01} initial mass function is assumed. For this study we define a dwarf galaxy as having a stellar mass $M_* \leq 3\times 10^9 \mathrm{M_\mathrm{\odot}}$. Throughout, the median of the mass probability distribution from the MPA-JHU catalogue is used as our mass value. Applying our dwarf galaxy threshold to this mass entry returns 65,461 galaxies.\\

MPA-JHU only consists of objects spectroscopically classified as galaxies by the SDSS  - those with absorption lines or emission lines with widths $ < 200 \mathrm{\ km\ s^{- 1} Mpc}^{- 1}$. To check if objects identified as broad-line AGN are being excluded from our dwarf galaxy sample we also searched the \cite{Shen11} SDSS DR7 quasar catalogue. It provides estimates of the central black hole masses using virial techniques. Here, we adopt scaling relations to estimate the total stellar masses of the host galaxies of these quasars and thus determine if any would enter our dwarf galaxy sample. 
We adopt a fiducial scaling between BH mass and bulge mass of $M_\mathrm{BH} = 0.002 M_\mathrm{bulge}$, and, conservatively, assume that the bulge mass corresponds to the total stellar mass of the host galaxy. Using our dwarf galaxy threshold, we find an upper black hole mass of $\leq 10^{6.8} \mathrm{M_\odot}$.
\cite{ReinesVolonteri15} relate a galaxy's stellar mass to the mass of central black hole in the following way, $\log(M_\mathrm{BH}/M_\mathrm{\odot}) = 7.45 + 1.05 \log(M_\mathrm{*}/10^{11} M_\mathrm{\odot})$. This relationship was derived using a sample of AGN; their masses were calculated using either broad line widths, reverberation mapping or dynamical methods. The \cite{ReinesVolonteri15} relationship predicts a upper black hole mass limit of $\leq 10^{5.9} \mathrm{M_\odot}$.
Whichever relationship is used, the outcome is the same: none of the \cite{Shen11} objects have black hole masses close to these values. Thus, MPA-JHU is the sole catalogue used in this paper. \\

In addition to the dwarf galaxy sample, we also define a sample of comparison galaxies whose mass limit is $M_\mathrm{*} > 1\times 10^{10} \  \mathrm{M_\odot}$ given the high incidence of AGN in higher mass galaxies. This high mass sample is only used as a comparison point to our dwarf galaxy sample from MPA-JHU and thus we do not add any broad-line objects from the Shen et al. catalogue.\\

As with the mass, SFR is measured using fits to the spectral energy distribution based on photometry from the full extent of the galaxy taken from the SDSS and GALEX (UV; \cite{Martin05}). MPA-JHU SFRs are calculated using the method outlined in \cite{Salim07}, who construct stellar population based on the \cite{BruzualCharlot03} population synthesis models and assume a \cite{Chabrier03} IMF.
Star formation histories were not single stellar populations, but the combination of an exponentially declining continuous star formation $\mathrm{\propto e^{-\gamma t}}$, with $0 \leq \gamma \leq 1\ \mathrm{Gyr}^{-1}$ uniformly distributed across this range, and with random starbursts superimposed.
These bursts were constructed such that the occurrence of a single event over the past 2 Gyr is 50\% and had a duration uniformly distributed in the 30-300 Myr range. 
Once constructed, each model is subjected to the \cite{CharlotFall00} dust attenuation model. 
The model SEDs at the redshift closest to the galaxy in question are, in turn, compared and their $\chi^2$ values evaluated. From this a probability distribution corresponding to a range of possible SFRs is produced. Throughout, the median of that SFR probability distribution is used as our SFR value.\\ 

Naturally, not all the observations are uniform in their quality so the sample had to be refined: 835,861 objects with a "good" photometric reliability flag were selected. This indicated valid results from the photometric fits applied by the MPA-JHU team thus producing a usable mass measurement. However, before we can identify their AGN content the X-ray data must first be considered.

\subsection{XMM-Newton}
The X-ray data used in this study comes from the 3XMM DR7 catalogue released in 2017 \citep{Rosen16}. It is based on 9,710 pointed observations with the XMM-Newton EPIC cameras which have a field of view $\approx 30$' and cover the energy range $\sim 0.2 -12$ keV.  DR7 contains $\sim 400,000$ unique X-ray sources based on 727,790 individual detections. Typical position errors for DR7 are $\approx 1.5$ arcsec. ($1\sigma$) and extend down to a flux limit of $\approx 10^{-15} \mathrm{erg\ cm^{-2}\ s^{-1}}$. 3XMM's angular resolution depends on the instrument in use: the PN detector has a FWHM of $\sim$ 6" and HEW of $\sim$ 16"; the MOS detectors have a FWHM of $\sim$ 5" and HEW of $\sim$ 15". For our study we use the unique source list rather than the individual detections. Our results are thus averaged over several individual observations for a significant number of sources. Using the data from this release, we performed a sky match to the  optical positions of the entire MPA-JHU catalogue using a search radius of 10" around every X-ray object; this yielded 3,440 matches. From this sample we summed fluxes in the 2 - 4.5 keV and 4.5 - 12 keV bands and converted them to luminosities in the 2 - 12 keV energy range using the MPA-JHU redshifts.  Since these objects are at such a low redshift, no rest-frame correction was applied.

\begin{figure}
	\centering
	    \includegraphics[width=\columnwidth]{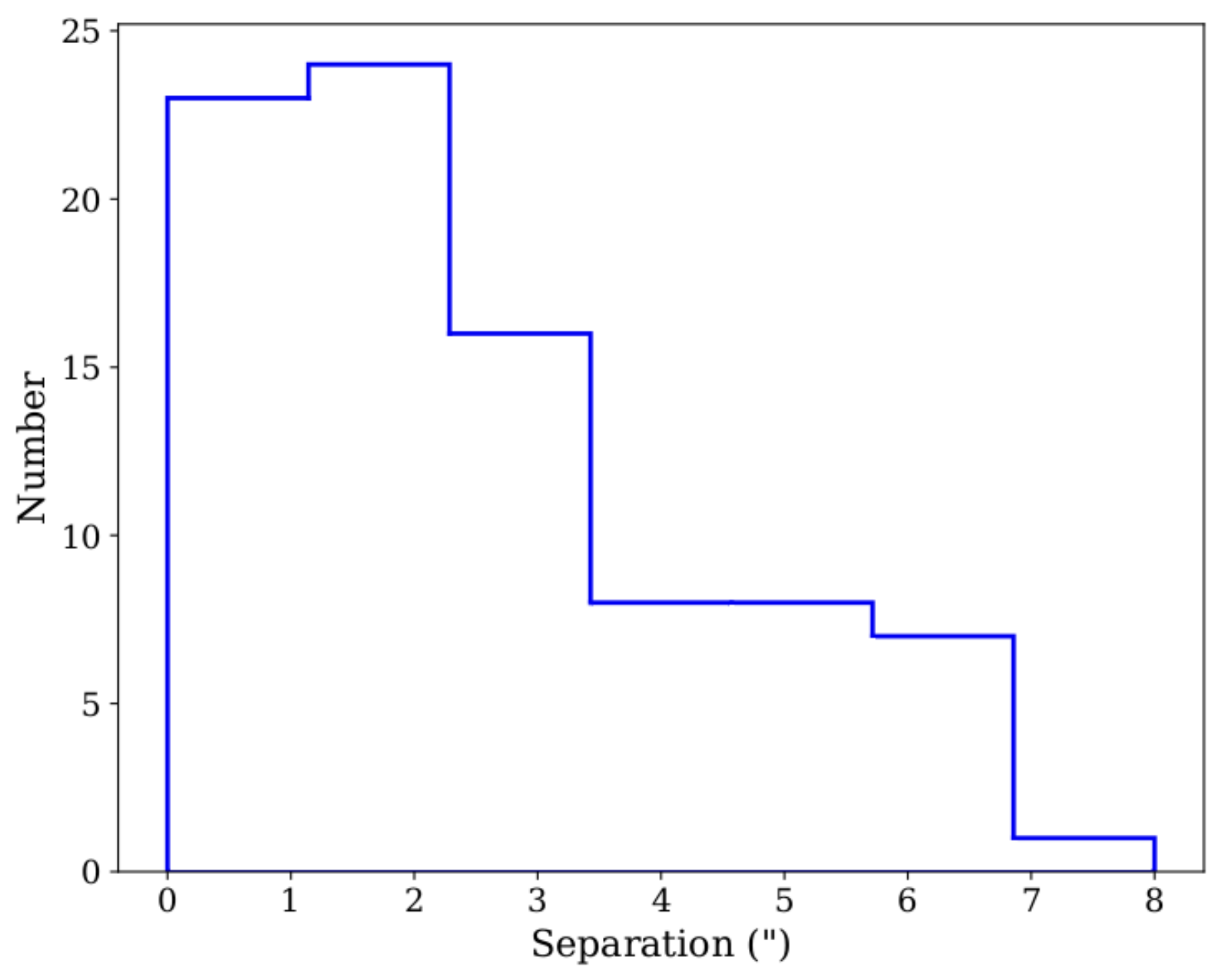}
	    \caption{Separation in arcseconds between the X-ray and optical signals for the 86 dwarf galaxies that meet the matching criteria outlined in section \ref{sec:matching}. 84\% of the signals in this sample are  matched within 5".}
	    \label{fig:sep}
\end{figure}

\begin{figure}
    \centering
    	\includegraphics[width=\columnwidth]{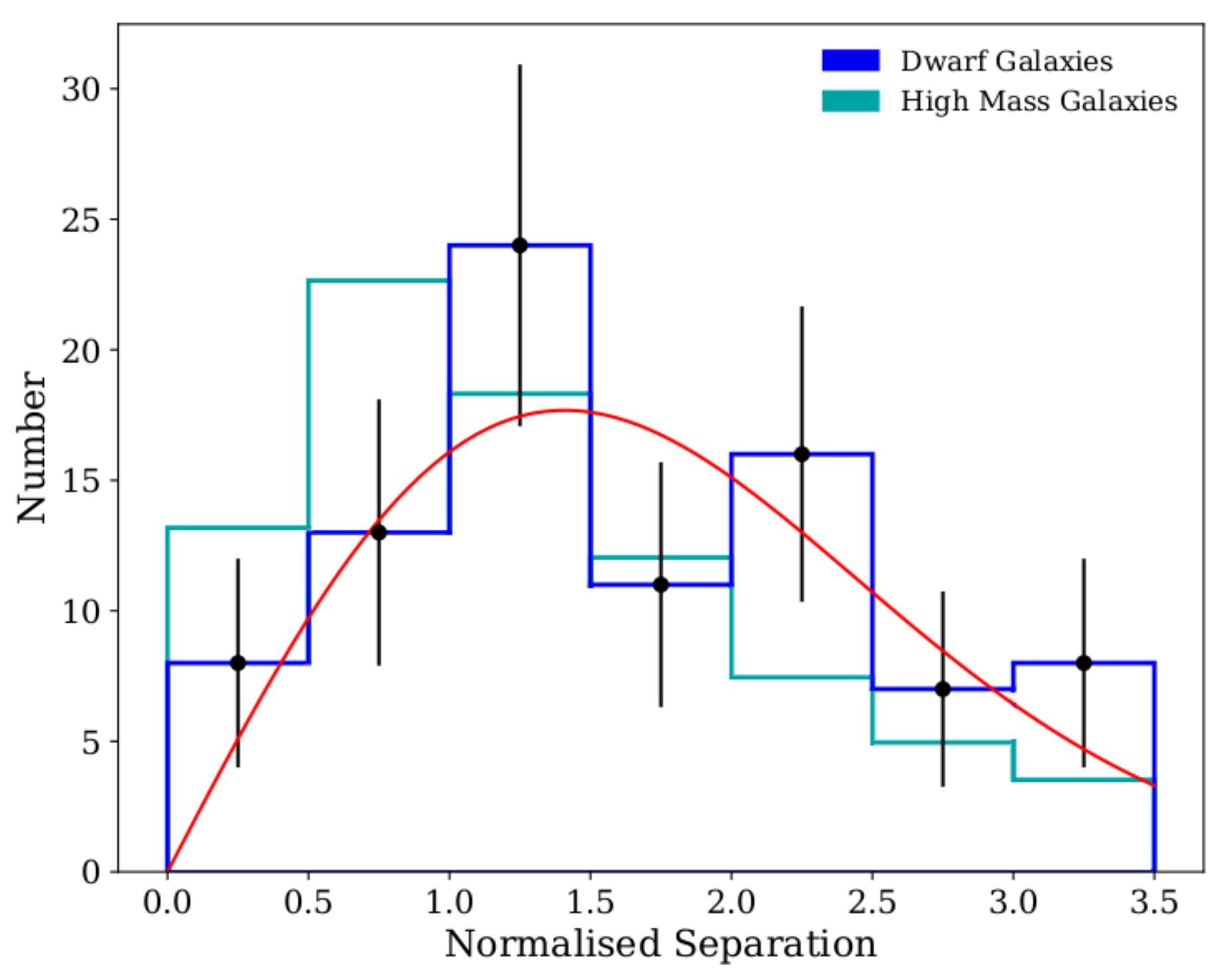}
        
        \caption{Position-error-normalised separation distribution (dark blue) for the 86 dwarf galaxies that met the matching criteria and other match checks described in section \ref{sec:matching}. Poisson errors are assumed in the histogram bins. The Rayleigh distribution fit (red) has a $\sigma = 1.4$ implying an underestimation in the XMM positional errors, a result that has been noted previously \citep{Watson09}. A sample of high mass galaxies (light blue) has undergone the same matching process and had its normalised separation scaled using the normalisation found from the Rayleigh fit.} 
               
        \label{fig:dataFitting}
\end{figure}

\subsection{Identifying Candidate AGN-hosting Dwarf Galaxies}
\label{sec:matching}

Using these X-ray active objects, a sample of possible AGN-hosting dwarf galaxies could be found. First, a set of more robust matching criteria were established to improve the confidence of association between the optical and X-ray objects. For each dwarf galaxy/X-ray pair we calculated the position-error-normalised separation, $x$, as follows,

\begin{equation}
   x = \frac{d_\mathrm{O,X}}{\Delta_\mathrm{X}}
\end{equation}

where $d_\mathrm{O,X}$ is the separation between the X-ray and optical signals, and $\Delta_\mathrm{X}$ is the error in the X-ray position. A dwarf galaxy was only considered a robust match if $x < 3.5$, giving a sample completeness of 99.8\%.
We also limited the extent of the X-ray source to less than 10" to ensure point-like emission consistent with an AGN \citep{Rosen16}. 
Applying this in conjunction with the $M_\mathrm{*} \leq 3 \times 10^9\ \mathrm{M_\odot}$ mass threshold yields 101 possible hosts. Applying the same matching criteria to the high mass sample yields 2,237 matches. \\

By imposing the dwarf galaxy mass limit we could have removed higher mass objects with a smaller separation to the X-ray signal, resulting in a poorer dwarf galaxy match being included. To check if any better matches to excluded objects exist, we uploaded the 101 X-ray co-ordinates to the SDSS SciServer and compared them to the full SDSS DR8 to find their nearest neighbour. 11 X-ray signals were found to have smaller separations to a higher mass galaxy removed by imposing the dwarf galaxy mass limit during the matching process. 

A visual assessment of the dwarf galaxy sample was performed. The optical images from the SDSS Finding Charts identified 3 sources that were extremely off-nuclear. If the photometry was constructed correctly for these objects then the refinement process outlined should have removed them, however this was not the case. We believed that in certain instances a galaxy that should not have passed our mass criterion was being broken into smaller sections and only having the MPA-JHU mass calculations applied to this small section. This was then matched to a nearby X-ray source, fooling our identification criteria into thinking it was a valid target. These 3 sources were removed from our sample leaving 87 dwarf galaxies with nuclear X-ray activity. This photometric fragmentation process has been observed in other, similar studies such as \cite{Sartori15}. 

A final assessment of the data found an object with extremely high redshift error. This was indicative of a poor fitting process leading us to doubt the accuracy of this object's data. It was removed leaving us with a final sample of 86 dwarf galaxies. Their distribution of separations between the X-ray and optical signals is shown in figure \ref{fig:sep}: 84\% of these dwarf galaxies match an X-ray object within 5", and the median matching radius is 2.2". A sample of dwarf galaxy images, with their X-ray signals and position error overlaid, can be found in appendix \ref{app:AGNImages}. 
Assuming a uniform distribution of the 65,461 dwarf galaxies in the SDSS area ($9274\ \mathrm{deg}^2$), we calculate a dwarf galaxy density of $7.05\ \mathrm{deg}^{-2}$. We then highlighted 131,736 X-ray sources within the SDSS area, from which we removed extended sources with the same criterion used to create the dwarf galaxy sample. We then calculate the sky area covered by these sources assuming we perform radial searches centred on the remaining 122,834 X-ray sources out to a radius of $3.5 \times \Delta_\mathrm{X}$. Multiplying the total search area, $0.91 \ \mathrm{deg}^2$, by the uniform dwarf galaxy density, tells us to expect about 6 false matches. Given a lot of these X-ray sources will already have well-defined counterparts, this value represents a conservative upper limit on the false match number. \\ 

As a final check of our position-error-normalised separation criterion, we plotted a distribution of this quantity for both samples of galaxies in figure \ref{fig:dataFitting}. The dark blue histogram shows the absolute values for the sample of 86 dwarf galaxies just identified, the light blue curve is the similarly-scaled high mass distribution. We perform a 1-D KS-test to compare this distribution to the expected distribution of XMM errors, given by the Rayleigh distribution. The KS-test returned a p-value of $10^{-4}$, indicating the observed distribution is not consistent with the expected Rayleigh distribution, given the fiducial positional uncertainties provide in the 3XMM catalogue. To investigate whether the observed distribution was better described by a skewed Rayleigh function, we fit it to this equation, 
\begin{equation}
f(x) = N \frac{x}{\sigma^2} e^{\frac{-x^2}{(2\sigma^2)}}
\end{equation}
where $x$ is the position-error-normalised separation. The normalisation constant, $N$ and $\sigma$ value were both free and we assumed Poisson errors for the number of objects per bin. Figure \ref{fig:dataFitting} shows the results of this fitting: it was found that the skewed Rayleigh distribution (red) had a $\sigma$ = 1.4, with a reduced $\chi^2$ of 1.90, implying an underestimate of the X-ray errors. It is also clear that both the dwarf galaxy and high mass distributions are consistent with this error underestimation. The underestimation of errors is a known issue in 3XMM, as noted in \citep{Watson09}. Given these positional uncertainties, our position-error-normalised separation criterion functions correctly, thus we expect our sources to be nuclear. 

\section{Analysing Dwarf Galaxy Emission}
\label{sec:emissionAnalysis}
Having identified 86 unique X-ray active dwarf galaxies by the matching process detailed in section \ref{sec:data}, it is necessary to break down their emission properties to determine the nature of the X-ray source.  In this section, we discuss the analysis carried out on the X-ray emission produced by the dwarf galaxies in this sample. 
This analysis has also been applied to a sample of 2,237 galaxies with $M_\mathrm{*} > 1\times 10^{10}\ \mathrm{M_\odot}$ as a comparison since these objects are more likely to harbour AGN. 

\subsection{Sample Properties}
\begin{figure}
	\centering
     	\includegraphics[width=\columnwidth]{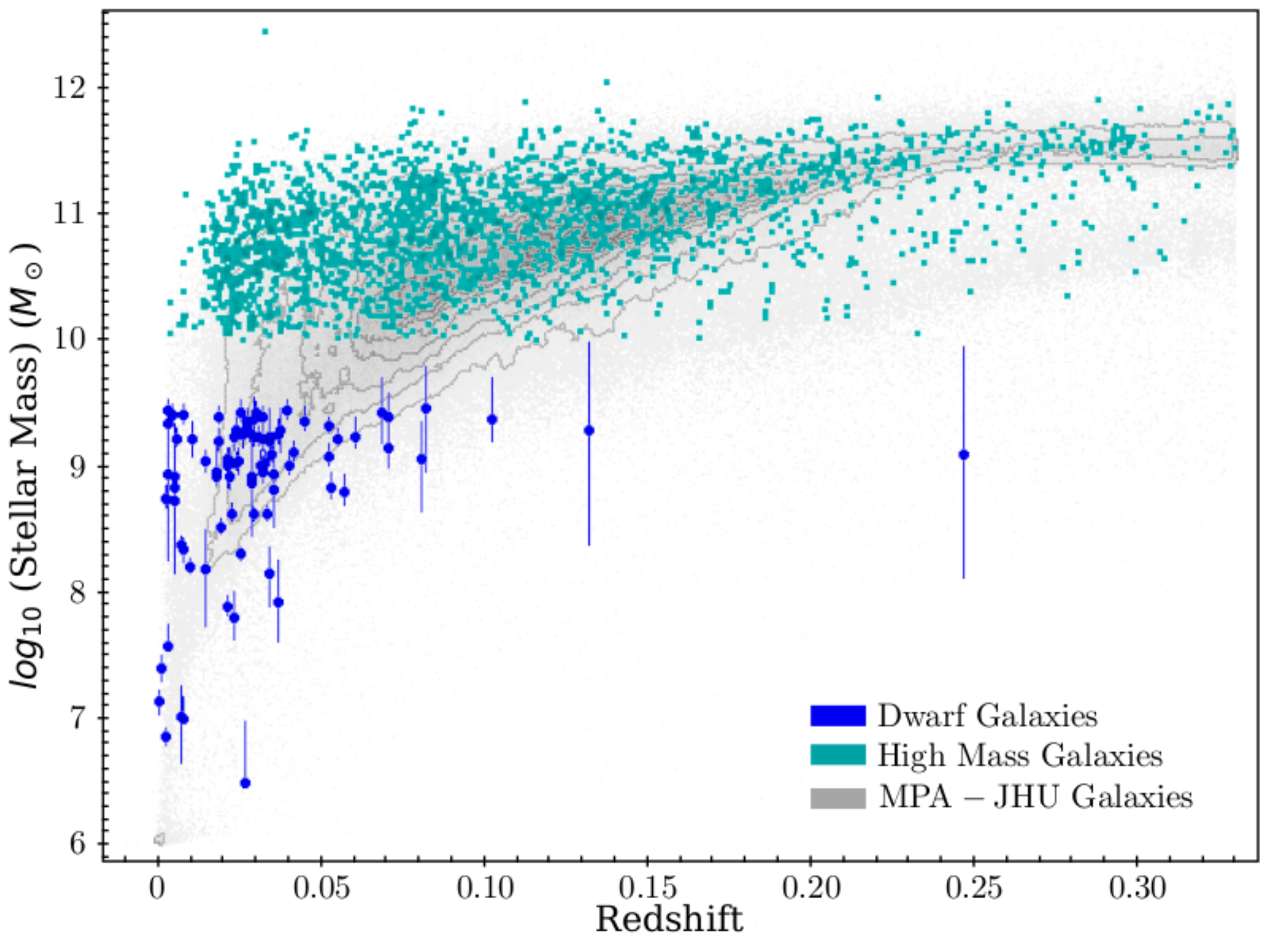}

		\caption{Mass against redshift for  X-ray active galaxies (grey) from the MPA-JHU catalogue. Highlighted are the points that make up the dwarf galaxy (dark blue) and high mass (light blue) samples. Section \ref{sec:emissionAnalysis} breaks down the X-ray emission properties of these samples.}
	\label{fig:MassZ}
\end{figure}

Figure \ref{fig:MassZ} shows the distribution of both samples of objects in mass and redshift. The dwarf galaxies, in dark blue, and high mass sample, in light blue, are plotted over the full sample of 3XMM X-ray detections within the SDSS area, in grey. Dwarf galaxies are found out to $z = 0.25$ and span a mass range of $3.08 \times 10^6 \mathrm{M_\odot}$ to $2.92 \times 10^9 \ \mathrm{M_\odot}$. Higher mass objects span a mass range $1.01 \times 10^{10}\ \mathrm{M_\odot}$ to $2.84 \times 10^{12}\ \mathrm{M_\odot}$ and are likely to host more luminous objects so can be detected at redshifts up to $z = 0.33$. \\ 

\begin{figure}
	\centering
  	\includegraphics[width=\columnwidth]{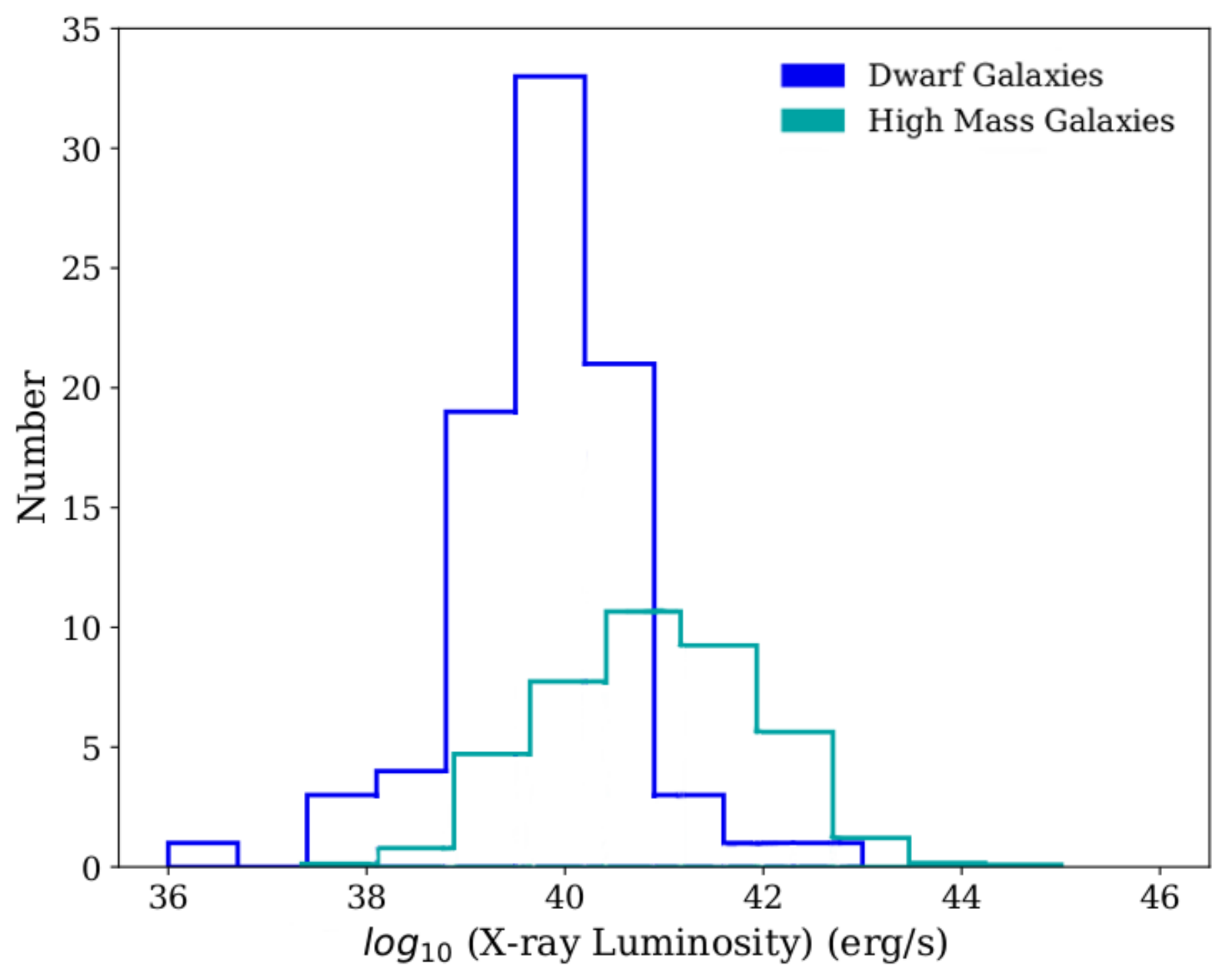}
\caption{X-ray luminosity distribution for the dwarf galaxy (dark blue) and high mass (light blue) samples. The high mass distribution has been arbitrarily scaled to allow for comparison.}

\label{fig:X_Ray_Lum}
\end{figure}

Figure \ref{fig:X_Ray_Lum} shows the distribution of X-ray luminosities for both samples. The dwarf galaxies, in dark blue, have observed X-ray luminosities of between $2.43 \times 10^{36}\ \mathrm{erg\ s^{-1}}$ and $5.35 \times 10^{42}\ \mathrm{erg\ s^{-1}}$; the modal group is between $5\times 10^{39}\ \mathrm{erg\ s^{-1}}$ and $1\times 10^{40}\ \mathrm{erg\ s^{-1}}$. Our dwarf galaxies are significantly less luminous than their high mass counterparts, shown in the arbitrarily-scaled light blue distribution, as expected. When identifying an AGN using only the X-ray luminosity, a threshold of $3 \times 10^{42}\ \mathrm{erg\ s^{-1}}$ is typically used; \citep{BrandtAlexander15} this would classify only 3 dwarf galaxies as containing an AGN. This criterion is clearly biased against the low luminosity AGN we expect in the dwarf galaxy sample so we must analyse the emission properties further.

\subsection{X-ray Contamination}
\label{sec:XRB}
Given most of the dwarf galaxy X-ray detections do not meet the suggested $3 \times 10^{42}\ \mathrm{erg\ s^{-1}}$ threshold, we need to assess other aspects of the emission. One possible source of X-ray contamination is the combined emission of unresolved X-ray binary (XRBs) populations within the galaxy. To estimate the contribution these objects make to a galaxy's X-ray emission, we used the relationship provided in \cite{Lehmer16} (see also \cite{Aird17}). It takes into account a galaxy's stellar mass ($M_*$), SFR and redshift (z) and takes the following form, 
\begin{equation}
L_{\mathrm{XRB}} = \alpha_0 (1+z)^\gamma M_\mathrm{*} + \beta_0 (1+z)^\delta SFR
\label{eq:Lehmer2016}
\end{equation}
where $\log_{10} (\alpha_0) = 29.37 \pm 0.15$, $\gamma = 2.03 \pm 0.60$, $\log_{10} (\beta_0)=39.28 \pm  0.03$ and $\delta = 1.31 \pm 0.13$ for 2 - 10 keV. \\

The low-mass X-ray binary (LMXB) contribution is correlated to the stellar mass, the high mass X-ray binary (HMXB) contribution to the SFR and the redshift dependence accounts for changes in metallicity and evolution of the XRB population. Using this relationship, we calculated the expected emission from every dwarf galaxy's binary population, $L_{\mathrm{XRB}}$. A total of 76 galaxies were found to have an observed luminosity that is greater than $L_{\mathrm{XRB}}$. Thus, there is a significant sample of X-ray emitting dwarf galaxies whose emission cannot be accounted for solely by their XRB population. \\

Hot gas in the interstellar medium can also produce X-rays, which could also potentially account for some of the emission. Its contribution can be estimated using the \cite{Mineo12b} relation,
\begin{equation}
L_{\mathrm{Gas}} = (8.3 \pm 0.1) \times 10^{38}\ SFR\ (\mathrm{M_\odot\ yr^{- 1}})
\end{equation}
We calculated the expected emission from the hot gas using \cite{Mineo12b} and added this to their $L_{\mathrm{XRB}}$. It was still useful to calculate this quantity despite the relatively low magnitude of this relation as all significant alternative X-ray sources needed to be considered. However, as expected all 76 objects which already have observed emission exceeding $L_{\mathrm{XRB}}$ also exceed the sum of $L_{\mathrm{XRB}}$ and $L_{\mathrm{Gas}}$. \\

\begin{figure}
	\centering
    	\includegraphics[width=\columnwidth]{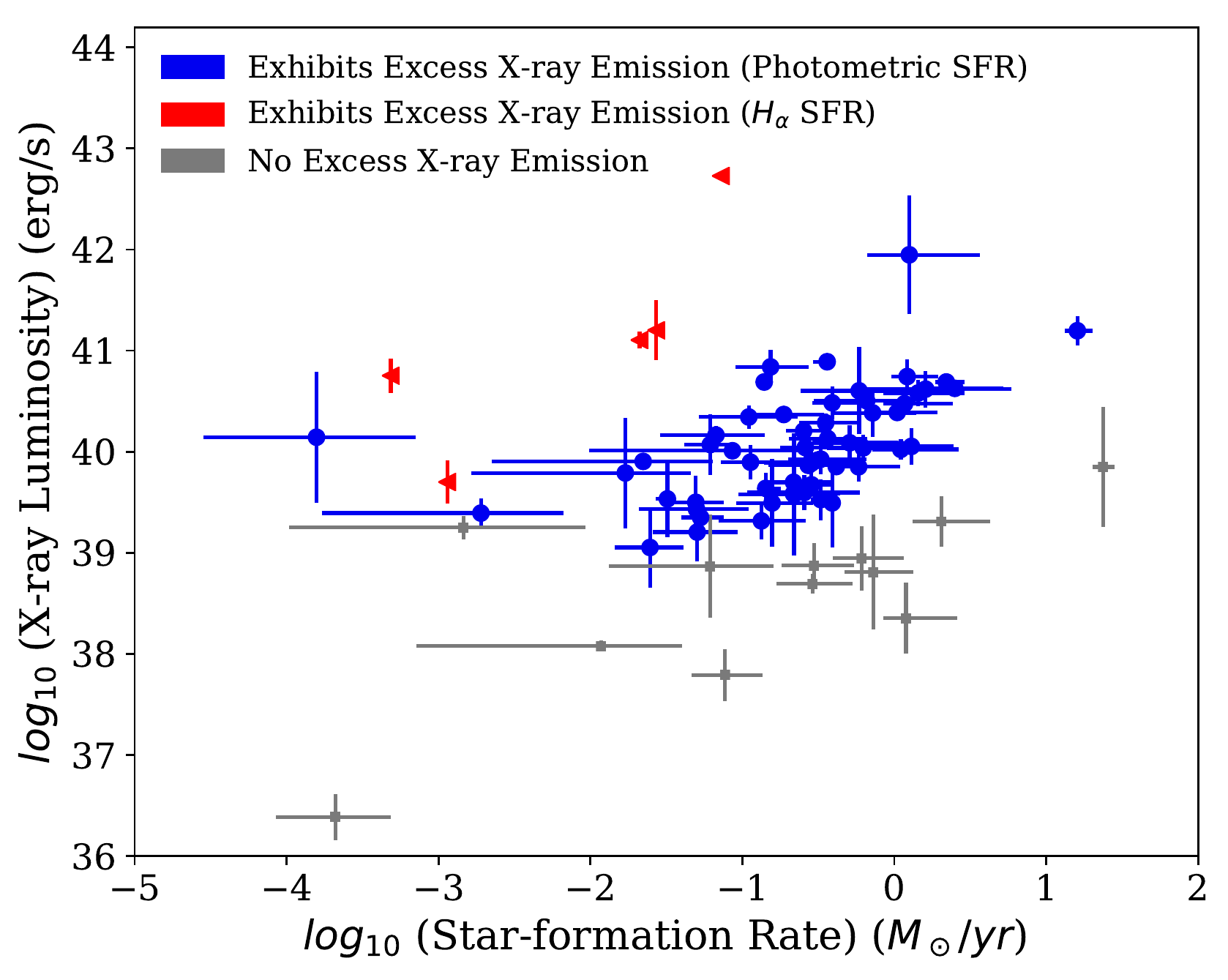}
    	
\caption{X-ray luminosity as a function of photometric star formation rate for the X-ray active dwarf galaxies. A total of 56 objects (blue) have observed X-ray emission which  exceeds the criterion of equation \protect \eqref{eq:excessCriterion}: possessing X-ray emission three times greater than the sum of $L_{\mathrm{XRB}}$, predicted X-ray emission from X-ray binary stars, and $L_{\mathrm{Gas}}$, predicted X-ray emission from hot gas. 5 objects (red) are those which meet or exceed the excess X-ray criterion but have their SFR calculated using the \protect \cite{KennicutEvans12} formalism and $H \alpha$ emission line. The grey points are those that did not meet the X-ray excess criterion.}
    
\label{fig:DG_Lx_SFR}
\end{figure}

Before we accepted this sample of 76 objects as AGN hosts, a level of significance needed to be applied to allow for potential uncertainties in the observed values of the relationships used. For this reason, dwarf galaxies which met or exceeded the following X-ray excess criterion,
\begin{equation}
\label{eq:excessCriterion}
\frac{L_{\mathrm{X,Obs}}}{L_{\mathrm{XRB}} + L_{\mathrm{Gas}}} \geq 3
\end{equation}
where $L_{\mathrm{X,Obs}}$ is the observed X-ray luminosity, were considered to have sufficiently excessive X-ray emission to potentially host an AGN. Figure \ref{fig:DG_Lx_SFR} shows the results of these calculations. A total of 61 objects highlighted in dark blue and red both have sufficiently excess emission to pass this criterion, however, they have their SFR measured in different ways. The 56 dark blue galaxies have SFRs calculated using the MPA-JHU method outlined in section \ref{sec:data}. The 5 red points, however, were flagged as having bad SFR fits by the MPA-JHU catalogue so instead we used the \cite{KennicutEvans12} formalism and the $\mathrm{H \alpha}$ emission line flux to give an SFR. Given the $\mathrm{H \alpha}$ emission line can be contaminated by AGN light, this SFR is a conservative upper limit. X-ray detections that do not meet the criterion given by equation \eqref{eq:excessCriterion} are plotted in grey. A breakdown of the observed and calculated properties for this sample of 61 galaxies can be found in appendix \ref{app:AGNData}. We also calculated the same quantities and applied the same criterion to the high mass sample which reduced it to 1,316 objects.

\subsection{Hardness Ratio}
The nature of the emission will also affect the shape of a galaxy's X-ray spectrum; this feature can be probed by measuring the hardness ratio (HR). HRs are defined between -1 and 1, with more positive values indicating harder X-ray emission, likely from an AGN, and more negative values indicating softer emission, likely from stellar processes. It is calculated using the counts from two different energy bands, A and B, 
\begin{equation}
    \label{eq:HR}
    HR = \frac{\mathrm{Band\ A - Band\ B}}{\mathrm{Band\ A + Band\ B}}
\end{equation}
We compare two HR bands: the soft band is the mean HR of the 0.5 - 1.0 keV and 1.0 to 2.0 keV bands; the hard band is the mean HR of the 1.0 to 2.0 keV and 2.0 - 4.5 keV bands. Analysing this quantity can help indicate whether the X-ray emission from our dwarf galaxy sample is coming from an AGN, rather than stellar processes. \\

\begin{figure*}
	\centering
	
    \includegraphics[width=\columnwidth]{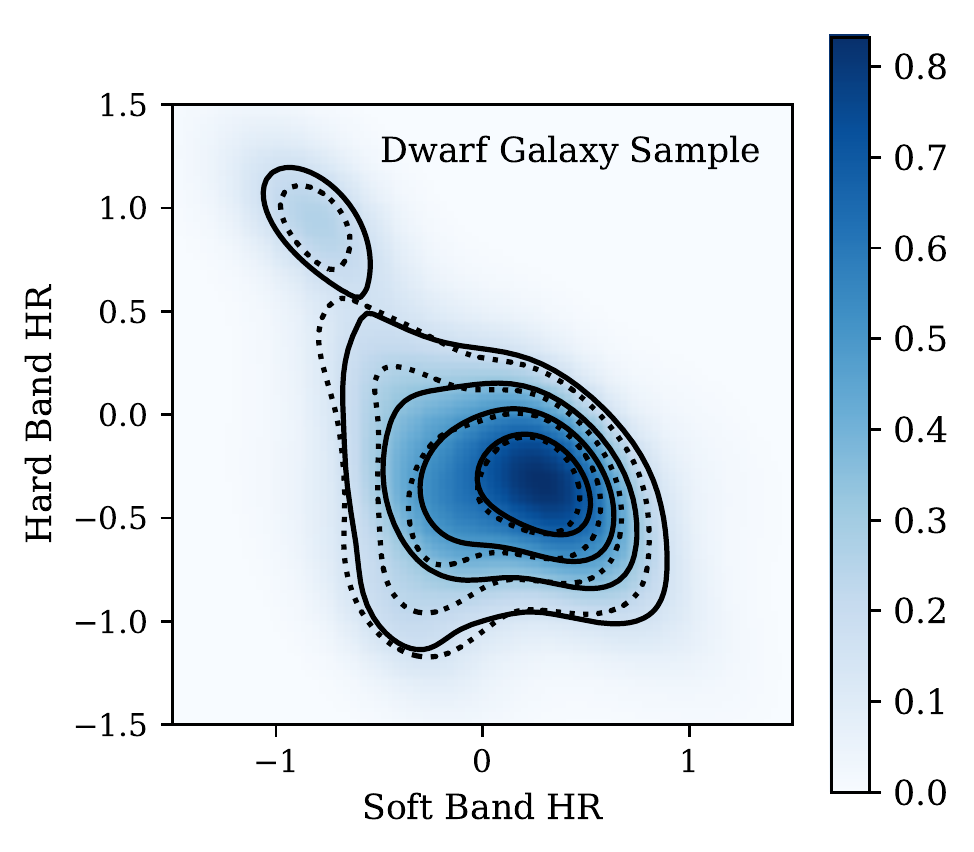}
    \includegraphics[width=\columnwidth]{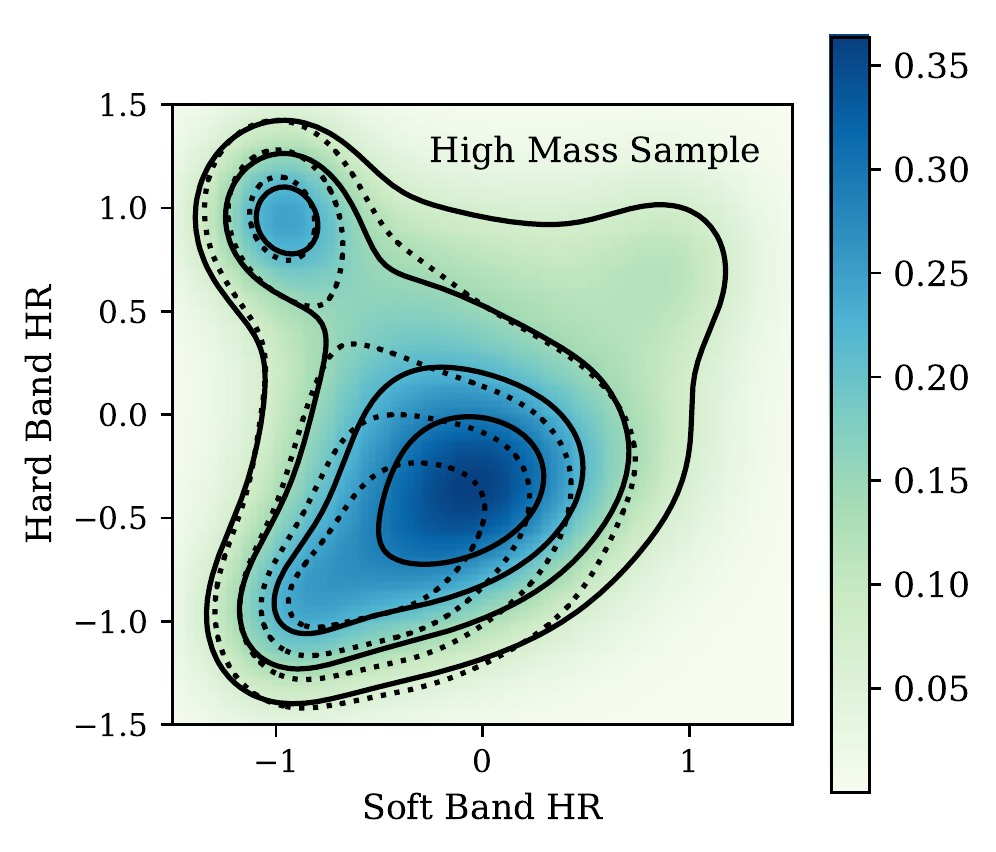}
    
        \caption{Error-weighted hardness ratios (HR) distribution for dwarf galaxy and high mass samples before (dotted line) and after (solid line) the application of the X-ray excess criterion in equation \eqref{eq:excessCriterion}. The soft band is the mean hardness ratio of the 0.5 - 1.0 keV and 1.0 to 2.0 keV bands; the hard band is the mean hardness ratio of the 1.0 to 2.0 keV band and 2.0 - 4.5 keV bands. Applying the X-ray excess criterion shows a positive shift in HRs, indicative of possible AGN activity. Encouragingly, both samples display similarities in the shape of their spectral distribution. The plots have been smoothed using a Gaussian kernel, whose width corresponds to the sample size and dimensions of the data: $\sim 0.5$ for the dwarf galaxies and $\sim 0.3$ for the high mass galaxies. In addition, the points are weighted by their errors. The contours encompass 80\%, 60\%, 40\% and 20\% of the samples in each plot.}
	\label{fig:HR}        
\end{figure*}

Figure \ref{fig:HR} shows the error-weighted distribution of hardness ratios for both our X-ray detected dwarf galaxy sample and high mass comparison sample before (dotted line) and after (solid line) applying the X-ray excess criterion in equation \eqref{eq:excessCriterion}. By overlaying the distributions in this way we can see equation \eqref{eq:excessCriterion} is working as intended: the emission in both distributions becomes, on average, harder. The contour lines can be seen to shift towards more positive values as objects with softer emission are removed. This effect is particularly pronounced in the high mass distribution, where a distinct group of harder emission objects had been isolated from the full high mass sample. Less strict attention was given to matching this sample so the high mass distribution, in the right-hand panel of figure \ref{fig:HR}, is more diffuse than that of the dwarf galaxies. However, higher mass galaxies generally have a larger number of confirmed AGN, thus we know this distribution will share some of their spectral properties. To determine whether the thresholded dwarf galaxy distribution shared any similarities with AGN spectra we performed two KS tests. Firstly, the thresholded dwarf galaxy distribution was compared with the full high mass distribution but they were found to be inconsistent, with a 2D 2 sample KS test producing a p-value of $10^{-3}$. When both the thresholded dwarf galaxy and thresholded high mass galaxies were compared, however, a 2D 2 sample KS test showed they are consistent at a $3\sigma$ confidence. Thus the emission from our thresholded dwarf galaxy sample shows characteristics of coming from AGN. \\

Spectra alone are not definitive in determining the source of emission. XRBs can also produce hard spectra, similar to an AGN. Plotting these spectral distributions next to each other and performing the KS tests, however, highlights similarities between the dwarf galaxy and high mass distributions which suggest the dwarf galaxy emission is dominated by AGN. \\

\section{BPT Classification}
\label{sec:BPT}
AGN can also produce signatures detectable in the visible part of the spectrum. The central accreting black hole ionises the surrounding gas causing various emission lines to come to prominence in the galaxy's spectrum. Other processes such as star formation can also ionise a galaxy's gas and dust but the associated radiation is much softer than from an AGN. The BPT diagram plots a pair of optical emission line ratios - $\frac{\mathrm{[OIII]}\ \lambda 5007}{\mathrm{H_\beta}}$ and $\frac{\mathrm{[NII]}\ \lambda 6583}{\mathrm{H_\alpha}}$ - against each other to try and distinguish the source of this ionising radiation \citep{BaldwinPhillipsTerlevich81}. Given the diagnostic's popularity in the field, we investigate whether the 61 AGN hosts we identified using X-ray selection techniques would also be found by the BPT diagram. Of the 61 AGN hosts in our sample, 53 had significant detections ($\frac{\mathrm{Line\ Flux}}{\mathrm{Line\ Flux\ Error}} > 3$) in each of the required emission lines so these were used in our analysis. \\

\begin{figure}
	\centering
    	\includegraphics[width=\columnwidth]{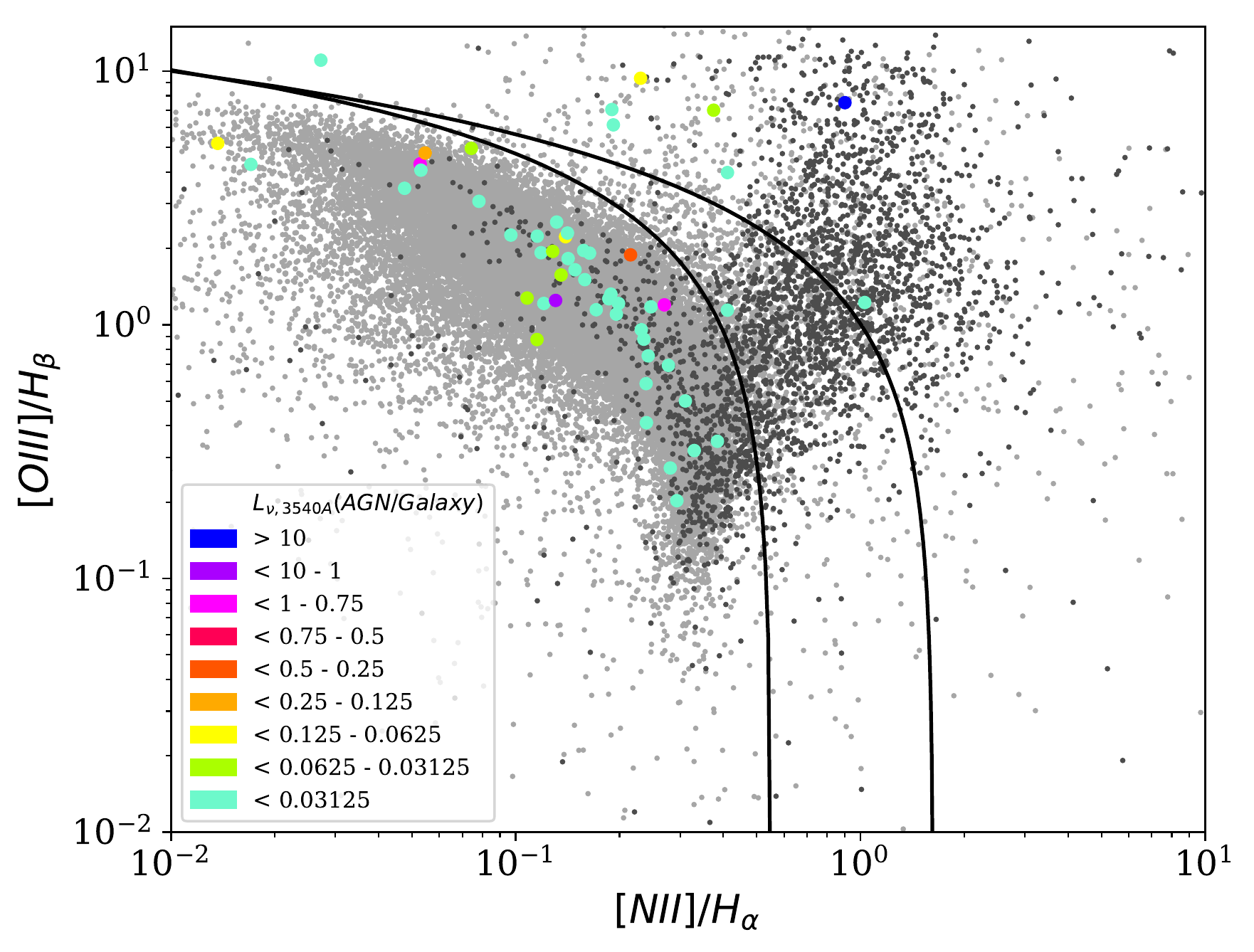}
        \caption{BPT diagram for various subsets of the MPA-JHU. The large, coloured points highlight the 53 dwarf galaxies identified as likely to contain an AGN with significant emission line detections. Their colour indicates the size of predicted optical AGN luminosity compared to that coming from the galaxy, both at $3450\dot{\mathrm{A}}$ - see section \ref{sec:BPT} for details on how this was calculated. The smaller, background points are defined as follows: the light grey points show all the low mass objects ($M_\mathrm{*} \leq 3 \times 10^9 \mathrm{M_\odot}$) found within the MPA-JHU highlighting a large distribution of points along the star-forming track; the dark grey points are the objects in the MPA-JHU that have an X-ray counterpart, largely dominated by the high mass AGN-like objects.}
	\label{fig:BPTcomparison}
\end{figure}

The results of the BPT analysis are shown in figure \ref{fig:BPTcomparison} with our AGN hosts  plotted as large, coloured points in the foreground. Black lines separate the AGN hosts into different classifications: objects with ionisation signatures predominately from AGN lie in the top-right, those dominated by star formation in the bottom-left, and those that have a composite spectra are in the central region \citep{Kewley01, Kauffmann03b}. Underneath these points, the BPT classification for the 62,703 galaxies in MPA-JHU with $M_\mathrm{*} \leq 3 \times 10^9\ \mathrm{M_\odot}$ and $z \leq 0.25$ are plotted in light grey. They are dominated by star-forming objects. The dark grey points show local galaxies in MPA-JHU with an X-ray counterpart within 10". These are much more spread out but will represent a range of objects including high mass AGN-hosting galaxies. \\

Figure \ref{fig:BPTcomparison} clearly shows that only 8 AGN hosts in our sample have been classified as an AGN by the BPT diagnostic, 1 has been classified as a composite and 44 have been classified as star-forming. Like other methods for identifying AGN, the BPT diagnostic has difficulty finding AGN in certain environments. \cite{Moran+02} notes that blue, star-forming galaxies present a challenging environment for the BPT diagnostic. The large amounts of star formation in these galaxies dominates the spectrum and hides emission from the lower luminosity AGN we expect to find.\\

To investigate whether star formation dominates optical emission in our AGN hosts we attempted to isolate emission from stellar and AGN processes at a single wavelength, $3450 \dot{\mathrm{A}}$ - the central wavelength of the SDSS u-band wavelength - to compare their magnitudes.

First, we consider the AGN emission. To ensure little contamination from stellar processes we used the X-ray emission to calculate the AGN's contribution to observed optical light. \cite{LussoRisaliti16} published the $\alpha_{\mathrm{OX}}$ relation which has the form, 
\begin{equation}
\log_{10}\ (L_{\mathrm{\nu,\ 2500\dot{\mathrm{A}}}}) = \frac{1}{0.6}\  (\log_{10}\ (L_{\mathrm{\nu,\ 2\ keV}}) - 7)
\label{eq:LS16}
\end{equation}
It relates the luminosity density at 2 keV, $L_{\mathrm{\nu,\ 2\ keV}}$ to that at $2500\dot{\mathrm{A}}$, $L_{\mathrm{\nu,\ 2500\dot{\mathrm{A}}}}$, from a sample of SDSS quasars. Assuming our dwarf galaxies follow the same relation it can provide a useful starting point in our efforts to estimate the effects of the AGN on optical observations.

To find the luminosity density at 2 keV, we calculated the geometric means of 3XMM bands 2 \& 3 and bands 4 \& 5, giving the luminosity densities at 1 keV and 5 keV respectively.  Using linear interpolation  between these values we calculated $L_{\mathrm{\nu,\ 2\ keV}}$  and used it in equation \eqref{eq:LS16} to find $L_{\mathrm{\nu,\ 2500\dot{\mathrm{A}}}}$. Translating $L_{\mathrm{\nu,\ 2500\dot{\mathrm{A}}}}$ to the required emission at 3450$\dot{\mathrm{A}}$ required the composite UV-optical quasar spectrum from \cite{VandenBerk01}. They model this region as a power law spectrum, $f_\mathrm{\nu} = c \nu^{-\alpha}$, where $\alpha = 0.44$. This required $L_{\mathrm{\nu,\ 2500\dot{\mathrm{A}}}}$ to be converted to flux density, $f_{\mathrm{\nu,\ 2500\dot{\mathrm{A}}}}$ which was used, first, to scale the power for each AGN host. From this an average spectrum was constructed and the flux density at $3450\ \dot{\mathrm{A}}$ extracted.

The galaxy emission was more straightforward to find, we simply converted the SDSS U-band magnitude to a flux density at $3450 \dot{\mathrm{A}}$. 
We then divided the predicted AGN emission at $3450 \dot{\mathrm{A}}$ by that from the galaxy and split each object into groups based on this value. These dictate the colours used for the AGN host points in figure \ref{fig:BPTcomparison}.\\

A total of 50 AGN hosts have predicted optical AGN contributions less than that from galaxy emission - 44 have AGN contributions > 25\% of the galaxy - the majority of which lie in the star-forming region. This helps confirm the idea that the AGN emission in this part of the spectrum is being hidden by star formation processes, causing the BPT diagnostic to mis-classify them. Of the 2 remaining AGN hosts with an optical AGN excess, only one lies in the AGN region. This host has the largest optical AGN excess in the sample, over 10 times that of the galaxy emission. The other lies within the star-forming region because it has a much smaller excess, only 1.08 times that of the galaxy. This is likely insufficient to produce emission lines of the appropriate proportions to move it into the AGN region. This further suggests that the BPT diagnostic is biased against identifying the low-luminosity AGN we expect to see in dwarf galaxies. \\

Some studies have shown that the \cite{LussoRisaliti16} relation has an increased  dispersion in the regime of dwarf galaxies. In addition, these galaxies may be X-ray weak relative to their UV emission, so these results could under-predict the AGN contribution \citep{Plotkin16, Baldassare17}. Despite this, we have still identified a significant number of X-ray selected AGN mis-identified as star-forming galaxies. This finding is consistent with a increasing body of work suggesting that optical spectroscopic measurements are insufficient to identify AGN in these environments \citep{Agostino18, Cann19}.

\section{Specific Black Hole Accretion Rate}
\label{sec:sBHAR}

\begin{figure} 
    \centering
		\includegraphics[width=\columnwidth]{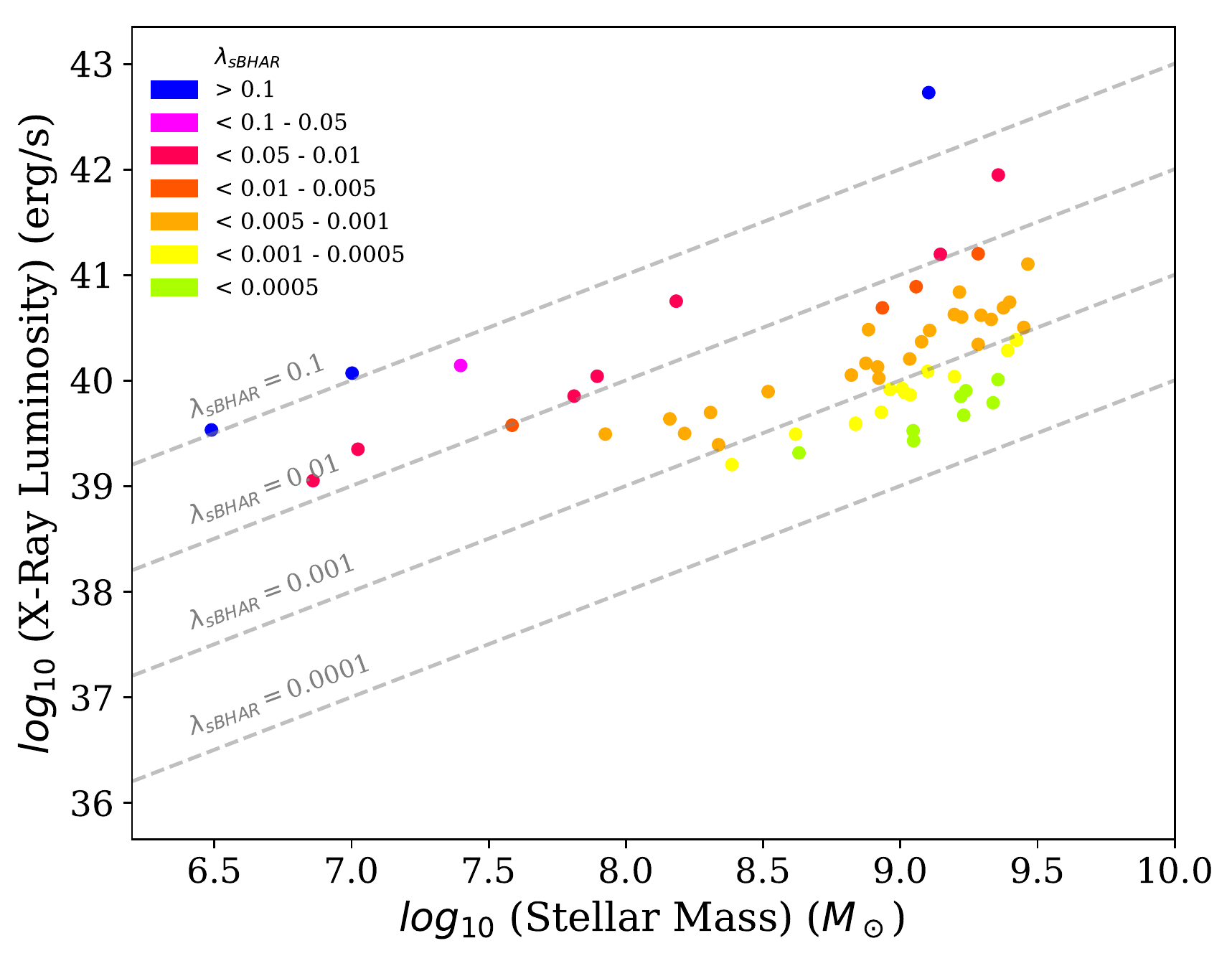}
    \caption{Figure showing the X-ray luminosity against galaxy mass for the 61 AGN hosts identified in section \ref{sec:emissionAnalysis}. The colour gives an indication of the accretion rate for the central black hole described in equation \eqref{eq:sBHAR}}.
    \label{fig:sBHAR}
\end{figure}

To investigate the activity of the central black holes powering our AGN we calculate their growth rates in terms of the specific black hole accretion rate (sBHAR), $\lambda_{\mathrm{sBHAR}}$. This compares the bolometric AGN luminosity of the galaxy with an estimate of the black hole's Eddington luminosity to give an indication of how efficiently the black hole is accreting. It is found using,
\begin{equation} \label{eq:sBHAR}
    \lambda_{\mathrm{sBHAR}} = \frac{25 L_{\mathrm{2-10 keV}}}{1.26 \times 10^{38} \times 0.002M_*} \approx \frac{L_{\mathrm{bol}}}{L_{\mathrm{Edd}}}
\end{equation}
and is taken from \cite{Aird12}. We assume that the black hole and stellar masses scale in the same way as their higher mass counterparts to get a sense of the black hole growth relative to the total galaxy mass. In order to clarify the validity of this assumption, we would need to compare this scaling relation to black hole masses but these are very difficult to accurately ascertain. Instead we looked at the morphologies of our AGN hosts to see if they are bulge dominated. Of the 36 AGN hosts found in the Galaxy Zoo DR1 \citep{Lintott11}, 33 have uncertain morphology and the remaining 3 are likely spirals. Despite this, we chose to cautiously continue using the relationship and present the results in figure \ref{fig:sBHAR}. It plots the mass of a particular host against its observed X-ray luminosity, the colour gives an indication of the sBHAR.  We have also plotted lines of constant sBHAR to give an idea of the typical mass and luminosity expected from these galaxies. 

The most common environment in our sample is a host of mass $\approx 10^{9 - 9.5}\ \mathrm{M_\odot}$ with an SMBH accreting at about 0.1\% of its Eddington luminosity. None of the SMBHs are accreting very efficiently; most of this sample have an sBHAR of less than 1\% of their Eddington luminosity with the most frequent accretion rate being around 0.1\%, with only one rising above 10\%. In the high mass region we see a wide range of sBHARs which is restricted as we move down the mass scale. This effect is likely due to the fact that we are missing objects with lower sBHAR at low masses as less efficiently accreting hosts will be more difficult to observe (due to their extremely low X-ray luminosities), and likely won't appear on the plot because of the threshold applied in equation  \eqref{eq:excessCriterion}.

\section{Completeness-corrected Luminosity and Accretion Rate Distributions}
\label{sec:UL}

Currently our sample is subject to significant observational biases as we preferentially identify the most luminous and actively accreting AGN, this gives us a skewed picture of the distribution of AGN in the wider galaxy population. In this section we attempt to correct these biases to try and understand how the probability of a galaxy hosting an AGN as a function of luminosity and accretion rate varies across the full dwarf galaxy population.\\

Firstly, we made sure that our sample of AGN were consistent with the bulk of the underlying galaxy population. Figure \ref{fig:MassZ} shows how the shape of the underlying galaxy distribution changes with increasing mass and that some of our observed AGN lie at a comparatively high redshift. To correct this we took galaxies of all masses in narrow bands of redshifts from MPA-JHU and worked out the mass which contained $\sim 90\%$ of the galaxies. Through this process we defined a mass completeness limit as a function of redshift. This completeness limit was applied to the observed AGN when we created 3 mass intervals: $9 - \log_{10}\ (M_\mathrm{*}/ M_\mathrm{\odot}) < 9.5$; $8 - \log_{10}\ (M_\mathrm{*}/M_\mathrm{\odot}) < 9$; $6.8 - \log_{10}\ (M_\mathrm{*}/M_\mathrm{\odot}) < 8$. As a result, our sample is also limited to $z \leq 0.06$.

In addition, our sample needs to be constrained so it only contains AGN with a statistically significant detection in the energy bands used in the X-ray completeness analysis. This analysis needs to be done to account for AGN that may have been missed due to the varying  sensitivity  of  3XMM. To do this we used Flix\footnote{Found at \href{https://www.ledas.ac.uk/flix/flix_dr7.html}{https://www.ledas.ac.uk/flix/flix\_dr7.html}}  \citep{Carrera07}, 3XMM's upper limits service. It provides an upper limit, flux estimate and associated error broken down by band and instrument for the whole of 3XMM.
We  use the upper limits and observed fluxes from band 8 in the PN camera; not only does this band cover the entire energy range of 3XMM but it also has the greatest number of AGN hosts that meet or exceed the 3XMM's detection threshold. Thus we restricted our observed AGN sample to only those with a detection likelihood  $> 6$ in this band. Taking this and our mass and redshift corrections into account leaves us with 28 AGN hosts in our statistical sample (these galaxies are marked with an asterisk in appendix \ref{app:AGNData}). \\

To account for the varying sensitivity, we need  to characterise the distribution  of X-ray detection upper limits  in the  region of  MPA-JHU  with coverage from 3XMM. There are 6,447 dwarf galaxies with matches to found within MPA-JHU and 3XMM, whose co-ordinates were uploaded to FLIX. It was able to produce X-ray flux limits at the co-ordinates of 4,331 dwarf galaxies - our parent sample. Using the dwarf galaxy's redshift, this flux limit could be turned into a luminosity. Once all the luminosity upper limits were recorded, they were converted into a cumulative histogram as a function of X-ray luminosity and normalised by the size of the parent sample. This gave us a distribution of the fraction of galaxies where an AGN could have been detected above a given X-ray luminosity, a distribution which will be referred to as the luminosity sensitivity function. We can use this information to correct the observed distributions of luminosities and account for the varying sensitivity of 3XMM, allowing us to recover estimates of the true underlying distribution of luminosities within our samples of dwarf galaxies.

\begin{figure*} 
    \centering
        \includegraphics[width=.9\paperwidth]{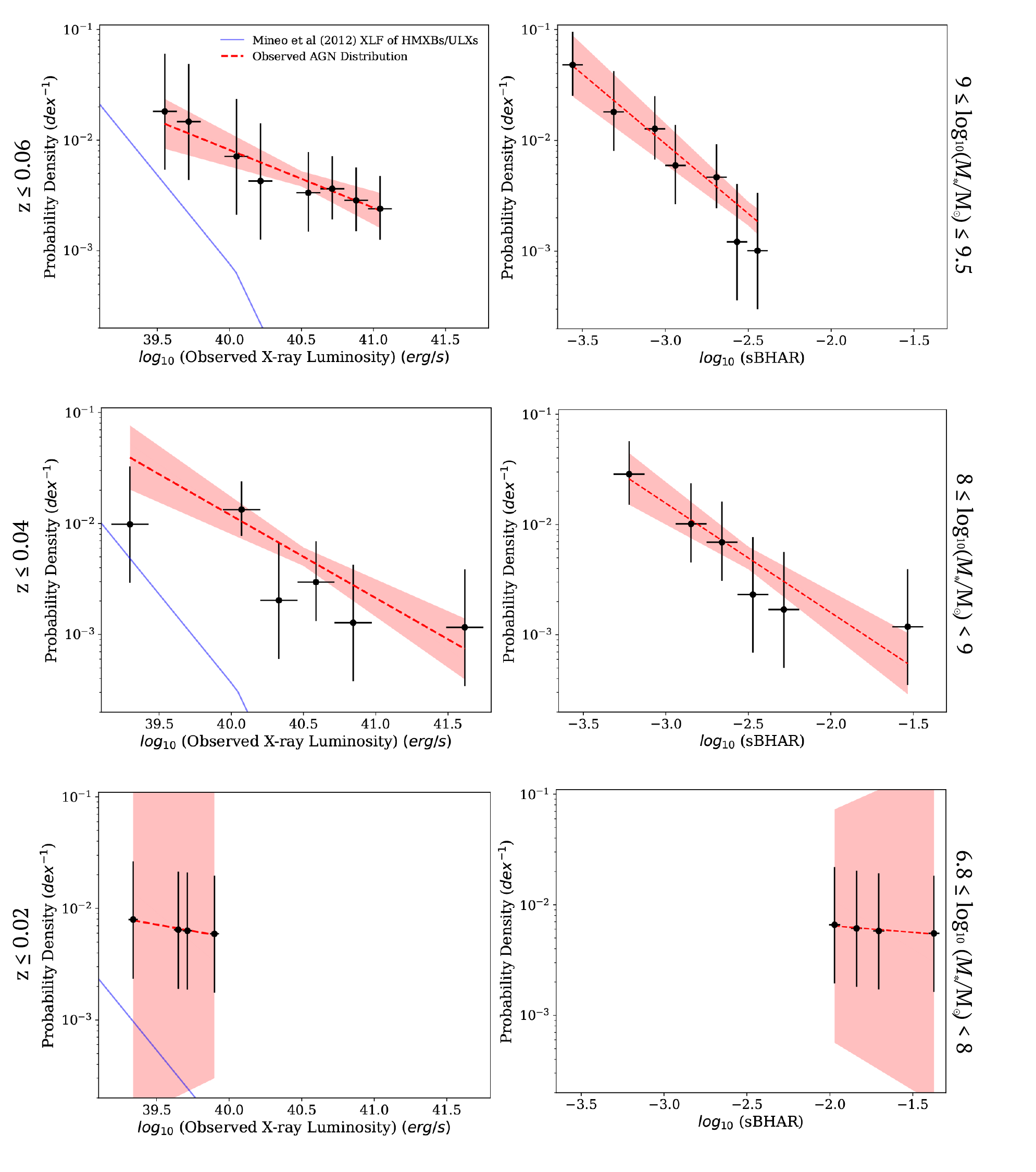}
    \caption{Distributions showing the probability of finding an AGN within the completeness corrected dwarf galaxy sample as a function of both observed X-ray luminosity (left-hand column) and specific black hole accretion rate (right-hand column). Power laws (dashed red line) were fit to each distribution, their uncertainty is also shown (pale red region). The redshift and mass applied to each row are shown. The blue line shows the expected average contribution from discrete, stellar-origin sources (HMXBs, ULXs) within our galaxy sample, based on the \protect \cite{Mineo12a} XLF scaled by the average SFR of the galaxies in our sample. It is plotted on each of the left-hand luminosity plots to rule out the possibility of a central ULX or group of ULXs being responsible for some of the low luminosity emission. See section \ref{sec:UL} for more details on how these plots were constructed.}
    \label{fig:UL}
\end{figure*}

To produce the probability distribution, the observed AGN in each mass interval were binned as a function of the observed X-ray luminosity. For each luminosity bin, we use the sensitivity function to determine the number of galaxies within our parent sample where the 3XMM data is sufficiently sensitive to detect an AGN of this luminosity. We then divided the total number of X-ray detections by the expected number of galaxies to provide an estimate the true probability of finding an AGN with such luminosities. This process produced the probability distributions seen in the left-hand column of figure \ref{fig:UL}. They show the probability of finding an AGN within nearby dwarf galaxies, in each mass and redshift interval, as a function of the observed X-ray luminosity.\\

Correcting the statistical sample of AGN with its corresponding luminosity sensitivity function has removed some of the observational bias described earlier. In contrast to the observed distribution shown in figure \ref{fig:X_Ray_Lum}, the probability of finding an AGN generally increases as we go to lower observed X-ray luminosities. Within each mass bin we can also see a similarly large spread of observed X-ray luminosities.

To produce the errors in the probability of hosting an AGN, we used the confidence limits equations presented in \cite{Gehrels86}. With this we estimated the error on the number of AGN in a given luminosity bin and then propagated them as fractional errors to our probability estimates.

Up until this point we have assumed all these objects are AGN, based on the low-mass \& high mass X-ray binary checks applied to our sample in section \ref{sec:emissionAnalysis}. Whilst the models used do consider the integrated X-ray emission from the galaxy up to around $10^{39} \mathrm{erg\ s^{-1}}$, an ultra luminous X-ray object (ULX) located in the galactic centre and emitting significantly more than the rest of the galaxy could potentially have been included in this sample. To check whether or not these objects had been included in our sample, we used the \cite{Mineo12a} X-ray luminosity function (XLF). It models the XRB populations, including ULXs, as a two-part power law normalised by the host galaxy's SFR. In each mass interval, the galaxies' XLFs were calculated and averaged to show how the average number of ULXs compared to the average number of AGN as a function of observed X-ray luminosity. As can be seen on the plots in the left-hand column of figure \ref{fig:UL}, the vast majority of the data points do not overlap with the ULX XLFs.\\ 
Whilst this large gap does exist, it is important to quantify how many ULXs we might expect in this sample. Stochastic star formation could produce a single, very luminous ULX which may account for some of the overlap we see in the lowest mass probability distributions. To calculate the number of ULXs we might expect, we folded the ULX luminosity function through the correction fractions extracted from the 3XMM sensitivity curve. This calculation suggests there are 1.26 individual ULX detections within our dwarf galaxy sample. Thus, the probability that 1 or more of our X-ray detections is in fact an ULX, rather than an AGN, is $\sim$72\% whilst the probability of 2 or more contaminants is $\sim$36\%, assuming a Poisson distribution.

It has also been noted that galaxies with lower than solar metallicities, like the dwarf galaxies being studied, have an enhanced HMXB population \citep{Brorby14}. Thus the \cite{Mineo12a} XLF could be underestimating the probability of finding a HMXB in this population of galaxies. \cite{Lehmer19} observes this enhancement in 4 dwarf galaxies of similar masses and metallicities to those in our sample. We follow the same procedure outlined earlier in the paragraph but use the normalisations outlined in \cite{Lehmer19} and observe some increase in the probability of finding a HMXB at any given X-ray luminosity. We have not shown this relationship on figure \ref{fig:UL} because the small sample size means the relationship will have large and undefined uncertainties and believe plotting the relationship would appear overly definitive. When the \cite{Lehmer19} relationship is plotted on the lowest galaxy mass panel we observe that the increase does overlap with the lower luminosity half of the fit line which suggests that some of these detections may not be AGN. However, when plotted in the higher host galaxy mass bins we see that the increase is insufficient to overlap with the probability distributions. Thus, we can be confident that the observed sample residing in the higher mass dwarf galaxies are AGN. \\

After confirming our sample were AGN we could confidently construct probability distributions for the sBHAR, as first considered in section \ref{sec:sBHAR}. Observed X-ray luminosity can be affected by a number of host galaxy properties so by doing this we can reduce the observational bias whereby a black hole growing at a given accretion rate in a low mass galaxy produces a similar observable X-ray luminosity when compared to a black hole with a lower accretion rate in a higher mass host galaxy. We can also confirm whether the results shown in figure \ref{fig:sBHAR} are consistent with the underlying population of AGN in dwarf galaxies. We repeated the upper limits correction process described earlier but instead binned the observed and upper limits data as a function of sBHAR. The results, plotted in the right-hand column of figure \ref{fig:UL}, show a similar dynamic range to the observed X-ray luminosities that varies significantly depending on the stellar mass bin. In addition as we move down the mass scale, the average sBHAR increases. This is consistent with the results in figure \ref{fig:sBHAR} as lower mass galaxies require a larger sBHAR to be observed. \\

A power law of the following form was fit to each of these plots,
\begin{equation}
    \label{eq:powerlaw}
       \log_{10}(p(X)) = A + k(\log_{10} (X) - x')
\end{equation}
where $p(X)$ is the probability of observing an AGN with corresponding X-axis quantity, $X$. Each equation is centred at $x' = $ 40.5 for luminosity and -2.5 for sBHAR, this is the median value for each quantity in the full statistical sample. The power laws are shown as a dashed red line allowing us to more clearly identify how the probability of finding an AGN in a dwarf galaxy, within a given mass and redshift, changes as a function of observed X-ray luminosity and sBHAR. We then performed a $\chi^2$ fit with a power law function, adopting the average of the \cite{Gehrels86} uncertainties as the error in each point. Parameter errors were then estimated by taking the square-root of the covariance matrix's diagonal. Using this, the uncertainty in the power laws could be highlighted as pink regions in figure \ref{fig:UL}.  

These power law fits reinforce the effect of the upper limits correction, giving us insight into the true extent of black hole activity across the dwarf galaxy population. We can confidently say that, in the higher mass plots (top and middle rows), the average number of AGN in dwarf galaxies as a function of both luminosity and sBHAR are well described by a power law. The luminosity power law in both these mass bins is distinct from the \cite{Mineo12a} XLF in both normalisation and index, highlighting the fact we have identified a distinct sample of AGN. Moreover, the luminosity power laws describing the high and middle mass samples are identical within the error regions. The sBHAR power laws do not have the same degree of similarity. However, they are very likely derived from a clean sample of AGN and their error regions are of a similar size to the luminosity power laws which suggests that sBHAR, like luminosity, could be a fundamental property of an AGN. This has been previously observed in \cite{Aird12} for a higher redshift and higher mass sample. 

A slight downward trend exists in the low mass bin but it is much more uncertain given only 4 AGN were identified in this region (compared to the 15 and 10 AGN found in the high and middle mass bins respectively). Thus each point is based on a single observation and leads to comparatively large errors.

Overall, we see that AGN in dwarf galaxies are detected emitting at a range of luminosities, driven by a correspondingly large range of accretion rates. For each mass bin, the average number of AGN increases with decreasing X-ray luminosity and sBHAR. This is consistent with other AGN population studies \citep{Georgakakis17,Aird18}, showing we expect AGN with lower luminosities and sBHARs to be much more numerous, despite our current inability to detect them.

\section{AGN Fraction as a Function of Host Galaxy Mass and Redshift}
\label{sec:fraction}
The fraction of galaxies that host an AGN within given mass and redshift regimes can be easily derived from the probability distributions shown in figure \ref{fig:UL}. In this section we use the luminosity probability distributions, in the left-hand column of figure \ref{fig:UL}, to calculate robust AGN fractions above fixed luminosity limits to provide a direct comparison with previous work.  \\

\begin{figure} 
    \centering
        \includegraphics[width=\columnwidth]{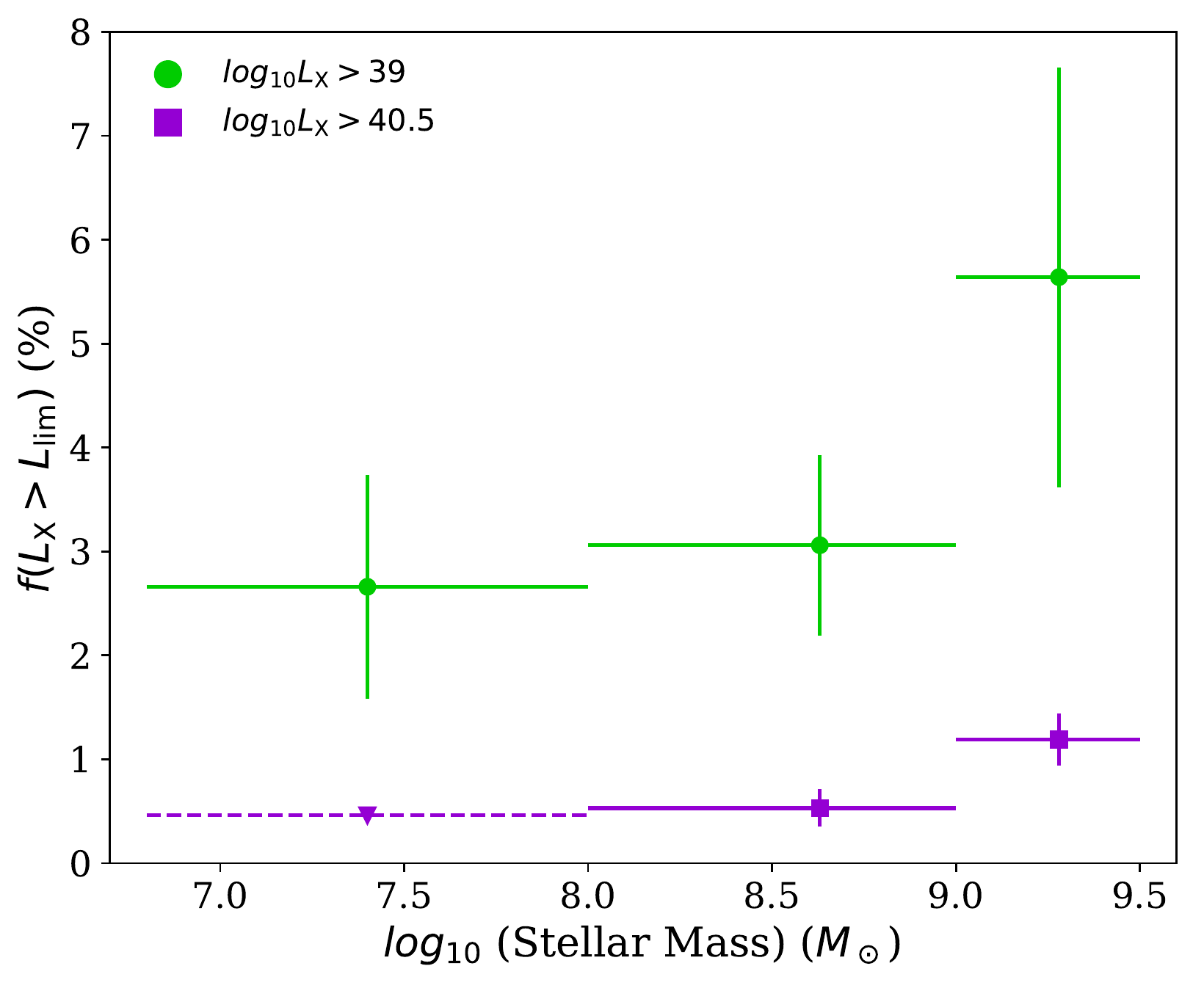}
    \caption{AGN fraction, $f(L_\mathrm{X} > L_\mathrm{lim})$, as a function of host galaxy mass for the full (green circles) and high luminosity (purple squares) samples as a function of host galaxy mass. AGN fraction increases with stellar mass in both luminosity regimes, however, the increase is more statistically significant for AGN with $\log_{10} (L_\mathrm{X}) > 40.5$.}
    \label{fig:Fraction_Mass}
\end{figure}

First, we consider how AGN fraction varies as a function of host galaxy mass. For each plot in the left-hand column of figure \ref{fig:UL} the probabilities were summed, converted into fractions, with the following equation,
\begin{equation}
    \label{eq:ProbSum}
    f(L_\mathrm{X} > L_\mathrm{min}) = \sum_{L_{\mathrm{min}}}^{42} p(\log_{10} (L_\mathrm{X})) \times \Delta \log_{10} (L_\mathrm{X})
\end{equation}
where $f(L_\mathrm{X} > L_\mathrm{lim})$ is the AGN fraction with X-ray luminosity $> L_\mathrm{min}$. The AGN fraction for the full range of X-ray luminosity ($L_\mathrm{X} > 10^{39} \mathrm{erg/s}$) are plotted in green on figure \ref{fig:Fraction_Mass} at the median mass for that sample. The process was then repeated with only the higher luminosity half of the observed AGN hosts ($L_\mathrm{X} > 10^{40.5} \mathrm{erg/s}$), and plotted as purple squares on the same figure. Overall we see an increase in AGN fraction with host galaxy mass. This is consistent with the expectation that black holes in higher mass galaxies are near ubiquitous, thus as as the incidence of black holes increases we would expect a corresponding increase in the incidence of AGN. We also saw in section \ref{sec:UL} that AGN with low accretion rates are more common, thus the likelihood that an AGN would produce detectable emission rises with host galaxy mass. This helps explain the large difference in AGN fraction for $\log_{10} (L_\mathrm{X}) > 40.5$. At a fixed host galaxy mass this measurement is highlighting the higher accretion rate end of the distribution. A statistically significant increase along the higher luminosity track would suggest increasing host galaxy mass is the dominant source of increasing AGN fraction in the higher luminosity regime. However, we cannot confirm this result for the lowest stellar masses in this regime as there were no AGN observed with these properties. The upper limit presented as a triangle is based on the assumption that a single AGN does actually exist; a black hole in a host of this mass would have to be accreting at an exceptionally high rate to produce detectable luminosities. 

The errors on the AGN fraction as a function of host galaxy mass, $f(L_\mathrm{X} > L_\mathrm{lim})$, are found by summing the errors on each data point in quadrature. For the full luminosity bin, $\log_{10} (L_\mathrm{X}) > 39$,  the errors are consistent with a constant fraction up to $M_\mathrm{*} \sim 10^{9.5} \mathrm{M_\odot}$. Perhaps host galaxy mass is not the only property that affect the incidence of AGN for the bulk of the population. \\

Thus, we also consider how the AGN fraction varies as a function of the host galaxy's redshift. \cite{Mezcua18} present some interesting work on this topic, looking at dwarf galaxies ($10^7 < M_\mathrm{*} < 3\times10^9 \mathrm{M_\odot}$) in the Chandra COSMOS-Legacy survey out to $z \sim 2.4$. Using similar techniques to those outlined in section \ref{sec:emissionAnalysis}, they identified a sample of 40 AGN which they use to study the evolution of the AGN fraction as a function of stellar mass, X-ray luminosity and redshift. Their sample allows them to measure the AGN fraction out to $z \sim 0.7$ and $\log_{10} (L_\mathrm{X}) \sim 42.4$; we have only observed one AGN that reaches luminosities in that range. 

\begin{figure} 
    \centering
        \includegraphics[width=\columnwidth]{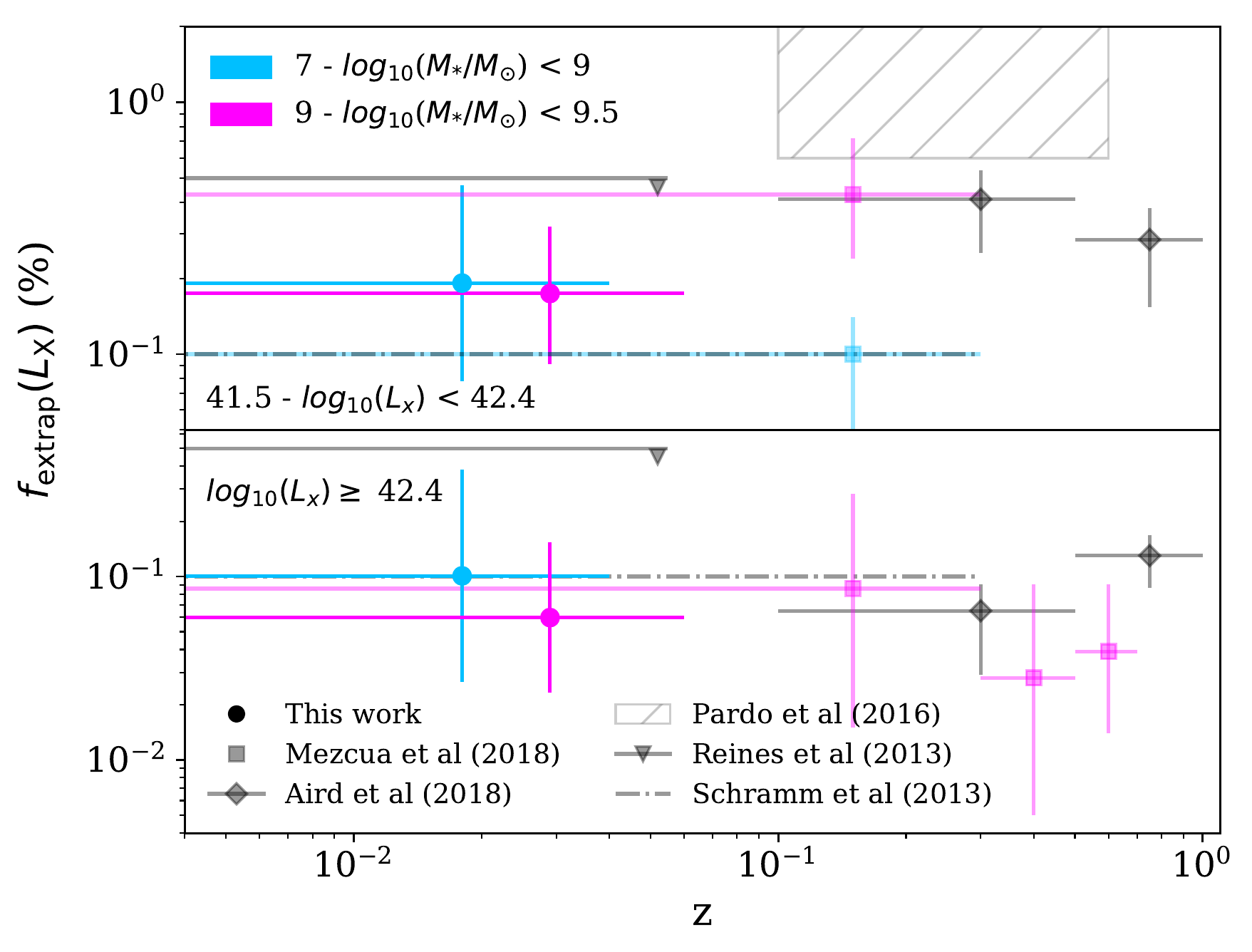}
    \caption{AGN fraction as a function of redshift for different observed X-ray luminosity ranges, adapted from \protect \cite{Mezcua18}. The circular points were derived from power law fits, within highlighted mass and corresponding 90\% complete redshift regimes discussed in section \ref{sec:UL}. Our results are consistent with most other work in the field.}
    \label{fig:Fraction_Z}
\end{figure}

By extrapolating the power law fit made in section \ref{sec:UL} we can estimate the AGN fraction at higher luminosities that are not directly probed by our study. We used the power law from the high mass bin ($9 < log_{10} ( M_\mathrm{*} / M_\mathrm{\odot}) < 9.5$) and re-ran the X-ray upper limits analysis for the $7 - \log_{10}\ (M_\mathrm{*} / M_\mathrm{\odot}) < 9$ used in \cite{Mezcua18}. We calculated the predicted AGN fraction as follows, 
\begin{equation}
    \label{eqn:ProbInt}
    f_\mathrm{extrap}(L_\mathrm{X}) = \int_{L_{\mathrm{min}}}^{L_{\mathrm{max}}} A + k (\log_{10} (L_\mathrm{X}) - 40.5)\ d \log_{10} (L_\mathrm{X})
\end{equation}
where $f_\mathrm{extrap}(L_\mathrm{X})$ is the predicted AGN fraction between the mass limits and luminosity limits shown in figure \ref{fig:Fraction_Z}. This figure shows our results plotted as circular points at the median redshift of each luminosity and mass bin, alongside the \cite{Mezcua18} data (square points), and from others studies as lines, regions and points. The errors in each of our AGN fractions is found by accounting for the $1\sigma$ uncertainty in the power law fit parameters.

Due to the limited redshift range of MPA-JHU, our points cover only the low redshift end of the axis, however in both panels they are consistent, within the errors, with the results from prior studies shown in figure \ref{fig:Fraction_Z}. 

For the lower luminosity AGN, in the upper panel, we find that our AGN fractions are consistent for both stellar mass bins showing no mass dependence. This contrasts with the results from \cite{Mezcua18}, taken at higher redshift. However, within the large uncertainties our data is consistent with Mezcua's, so we don't find any evidence for a mass dependence in lower luminosity AGN. 

Lower fractions in the bottom panel of figure \ref{fig:Fraction_Z} show that higher luminosity AGN are less abundant than their dimmer counterparts. This is consistent with earlier findings suggesting that AGN are more numerous at lower observed X-ray luminosity. \cite{Mezcua18} suggested that the AGN fraction may decline with increasing redshift, although given the large uncertainties, their measurements are consistent with a constant AGN fraction out to $z = 0.7$. Our result, in this mass and redshift bin, provides an estimate at low redshift, consistent with the Mezcua results at higher redshifts. Our results, therefore, indicate that the AGN fraction in dwarf galaxies is constant with increasing redshift.

\section{Summary and Conclusions}
This study has rigorously matched a sample of dwarf galaxies from the MPA-JHU catalogue to their central X-ray counterpart from 3XMM. We then predicted the contribution that XRBs and hot gas may make to this X-ray emission based on each galaxy's mass and SFR. Any galaxies that had observed emission three times greater than what was predicted were considered likely to be host an AGN. We showed that placing this threshold, likely isolated the hardest X-ray emission in this sample. Off-nuclear emission cannot be completely ruled out given the resolution of 3XMM, however recent simulations by \cite{Bellovary19} and observations in \cite{Reines19} suggest that such a detection could still be an AGN. Thus, we can confidently say we have identified 61 dwarf galaxy AGN hosts. Of these AGN, 21 have been previously identified (see appendix \ref{app:AGNData} for details).\\

We then performed BPT analysis on the 53 AGN hosts with significant detections in the required emission lines to see if their optical characteristics matched their X-ray classification. We found that the vast majority of these galaxies were classified as star-forming. Our result adds to a growing body of evidence suggesting that optical selection methods may miss AGN, particularly those residing in dwarf galaxies. To investigate if star formation dominated the optical component of AGN emission, we took the X-ray emission - dominated by the AGN - and translated it down to the SDSS U-band, to find out the AGN's contribution to the host galaxy's optical emission which is compared to the observed u-band flux. All but one of the X-ray selected AGN classified as star-forming had their optical emission dominated by the galaxy. This finding shows that star formation can confuse the results from BPT diagnostics and hide the signatures of AGN.\\

Next we investigated the activity of the central SMBHs powering our AGN by calculating their accretion rates. We found that the SMBHs span a wide range of accretion rates but that none are accreting very efficiently. The most common environment in our sample is a host of mass $\sim  10^{9 - 9.5} M_\odot$ with an SMBH accreting at about $0.1\%$ of its Eddington luminosity. The most active SMBH has an accretion rate of $\sim 10\%$ of its Eddington luminosity.\\

Finally, we attempted to correct our sample to account for AGN that could have been missed due to the varying sensitivity of 3XMM. To do this, the observed AGN sample needed to be reduced so it matched the underlying galaxy distribution and had significant detections in the appropriate 3XMM band. Thus our observed sample was reduced from 61 to statistically robust sample of 29 AGN, which were then split into 3 mass and redshift bins. To correct the observed distribution, we determined the upper X-ray flux limit at the positions of the 4,331 dwarf galaxies in MPA-JHU that lie within the 3XMM footprint. We use these flux upper limits to determine a sensitivity function that allows us to correct our observed distributions of AGN luminosities for incompleteness and recover the true probability distribution functions of AGN luminosities and specific accretion rates within the dwarf galaxy population. As a final check, the luminosity probability distributions were compared to the \cite{Mineo12a} XLF of HMXBs and ULXs. They were found to lie above the XLF indicating that we are identifying a distinct population of AGN within dwarf galaxies and are not significantly contaminated by the detection of individual, bright ULXs within the galaxy. 
The probability distributions show that AGN in dwarf galaxies have a wide range of activity, with the probability of identifying an AGN being well described by a power law. AGN are more numerous at lower X-ray luminosities and sBHARs.\\

We used our robust measurements of the probability of hosting an AGN as a function of X-ray luminosity to determine how the incidence of AGN varies as a function of other galaxy properties. We find evidence that the fraction of galaxies with an AGN above a luminosity limit of $L_\mathrm{X} > 10^{39}$ erg/s increases as a function of stellar mass, rising from $\sim 2.7$\% at $M_\mathrm{*} \sim 10^{7.4} \mathrm{M_\mathrm{\odot}}$ to $\sim 6$\% at $M_\mathrm{*} \sim 10^{9.5} M_\mathrm{\odot}$. We extrapolate our measurements to higher luminosity thresholds and compare to higher redshift measurements from \cite{Mezcua18}, finding no evidence for any evolution in the AGN fraction in dwarf galaxies between $z \sim 0.7$ and $z \sim 0.03$ probed by our study.\\

In conclusion, we have shown that AGN with a broad range of accretion rates are found across the dwarf galaxy population. Our study shows that many of these AGN will be missed by the standard optical selection tools but are revealed by the careful analysis of X-ray observations of sufficient depth. Thus, accreting central massive black holes appear to be a common feature even in at the lowest galaxy masses. Our measurements quantify the AGN fraction, providing lower limits on the incidence of massive black holes in this low-mass regime and thus crucial constraints on the physical mechanisms that determine their  formation and subsequent growth.

\section*{Acknowledgements}
We thank the referee for their helpful comments.
KB acknowledges funding from a STFC PhD studentship. JA acknowledges support from an STFC Ernest Rutherford Fellowship, grant code: ST/P004172/1.

This research has made use of data obtained from the 3XMM XMM-Newton serendipitous source catalogue compiled by the 10 institutes of the XMM-Newton Survey Science Centre selected by ESA.

In addition, this research made use of Astropy,\footnote{http://www.astropy.org} a community-developed core Python package for Astronomy \citep{astropy:2013, astropy:2018}.

Funding for SDSS-III has been provided by the Alfred P. Sloan Foundation, the Participating Institutions, the National Science Foundation, and the U.S. Department of Energy Office of Science. The SDSS-III web site is http://www.sdss3.org/.

\input{ms.bbl}

\appendix 
\section{Images illustrating cross-matching procedure}
\label{app:AGNImages}
Figure A1 provides examples from our dwarf galaxy AGN sample and illustrates results of the cross-matching process. The images are ordered by increasing sky separation between the optical centre of the galaxy (black cross) and the X-ray signal (magenta cross). In the top left panel, the X-ray signal is 0.3" away from the optical centre of the galaxy, and it sits well within the X-ray position error (solid circle). In contrast, the optical centre of the galaxy in the bottom right panel sits at the edge of the 3.5 $\times$ position error (dashed circle) as it lies 6.8" away from the X-ray signal.

\begin{figure*}
    \centering
    \includegraphics[width=.9\paperwidth]{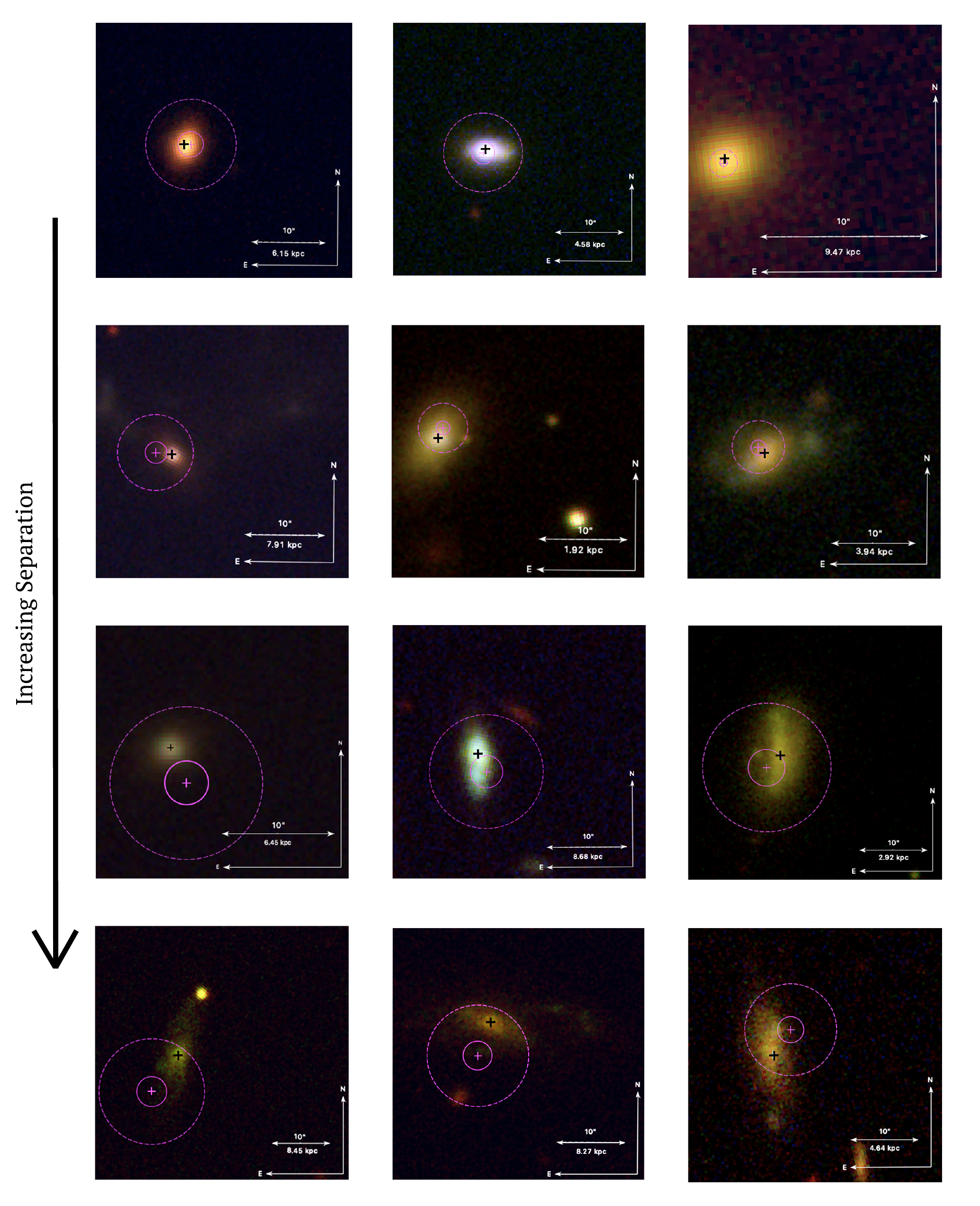}
    \caption{A sample of 12 showing the positions of the galaxy (black cross) and X-ray source (magenta cross) surrounded associated position error (solid circle) and 3.5 $\times$ position error (dashed circle). These images are ordered by increasing sky separation between these signals.}
\end{figure*}

\section{List of AGN Candidates}
\label{app:AGNData}
Table B1 contains data for the 61 dwarf galaxies that we identify as hosting X-ray AGN. The table is split into two sections: first are the AGN with a valid photometric SFR, followed by the AGN whose SFR was calculated using the $H_\mathrm{\alpha}$ luminosity and \cite{KennicutEvans12} method. 
Each galaxy's co-ordinates have also been uploaded to SIMBAD and NED to check if these AGN have been previously detected. 
The columns are defined as follows:

\begin{itemize}
    \item (1) Common name of galaxy taken from SIMBAD and NED. Galaxies marked with an asterisk are those which made up the statistical sample (see section \ref{sec:UL} for more details). 
    \item (2) A letter in this column indicates that the AGN has been identified previously. Here follows a list of the studies which have previously identified our AGN, the letters used to represent them and the wavelength range within which their study focuses: 
    \begin{itemize}
        \item a: \cite{VeronCetty10} (Optical, X-ray \& Radio)
        \item b: \cite{ReinesGreenGeha13} (Optical)
        \item c: \cite{Lemons15} (X-ray)
        \item d: \cite{Sartori15} (Optical \& IR)
        \item e: \cite{Sun15} (Optical)
        \item f: \cite{Baldassare17} (X-ray \& UV)
        \item g: \cite{Kawasaki17} (Optical)
        \item h: \cite{Marleau17} (IR)
        \item i: \cite{Nucita17} (X-ray)
    \end{itemize}
    
     \item (3) \& (4) Galaxy co-ordinates taken from MPA-JHU (based on SDSS DR8).
    \item (5) \& (6) X-ray co-ordinates taken from 3XMM DR7.
    \item (7) Separation between the optical and X-ray co-ordinates in arcseconds.
    \item (8) Total stellar mass of galaxy and associated errors in units of $M_\mathrm{\odot}$. The median error for the upper/lower error on the mass is 0.1/0.08 dex.
    \item (9) SFR and associated errors in units of $M_\mathrm{\odot}\ yr^{-1}$. The median error for the upper/lower error on the SFR is 0.27/0.21 dex.
    \item (10) Observed 2-12keV X-ray luminosity and associated errors, taken from 3XMM, in units of erg/s.
    \item (11) Predicted 2-12keV X-ray luminosity ($L_\mathrm{XRB} + L_\mathrm{Gas}$; see section \ref{sec:emissionAnalysis} for more information) in units of erg/s.

\end{itemize}

\onecolumn
\begin{small}
\begin{landscape}
\begin{longtable}{lcccccccccc}

\hline	
\rowcolor{white}
& Previously &\multicolumn{2}{c}{Optical (${}^o$)}&\multicolumn{2}{c}{X-ray (${}^o$)}&&&&&\\
\rowcolor{white}
Name & found? & RA & Dec &RA & Dec & Separation ($"$) & Mass ($M_\mathrm{\odot}$) & SFR ($M_\mathrm{\odot}\ yr^{-1}$) & $L_\mathrm{Obs}$ (erg/s) & $L_\mathrm{Pred}$ (erg/s)\\
\rowcolor{white}
(1)& (2) & (3) &(4) & (5) & (6) & (7) & (8) & (9) & (10) & (11)\\
\hline
\endfirsthead

\hline	
\rowcolor{white}
& Previously &\multicolumn{2}{c}{Optical (${}^o$)}&\multicolumn{2}{c}{X-ray (${}^o$)}&&&&&\\
\rowcolor{white}
Name & found? & RA & Dec &RA & Dec & Separation ($"$) & Mass ($M_\mathrm{\odot}$) & SFR ($M_\mathrm{\odot}\ yr^{-1}$) & $L_\mathrm{Obs}$ (erg/s) & $L_\mathrm{Pred}$ (erg/s)\\
\rowcolor{white}
(1)& (2) & (3) &(4) & (5) & (6) & (7) & (8) & (9) & (10) & (11)\\
\hline
\endhead

\hline
\rowcolor{white}
\multicolumn{8}{l}{\textbf{Table B1:} Data for all 61 dwarf galaxies believed to be strong candidates for hosting an AGN.}&\multicolumn{3}{r}{continued on next page...}\\
\hline
\endfoot
   
\hline
\rowcolor{white}
\multicolumn{11}{l}{\textbf{Table B1:} Data for all 61 dwarf galaxies believed to be strong candidates for hosting an AGN. }\\
\hline
\endlastfoot    

\rowcolor{Gray}
SDSS J011523.96+003808.7 *&&18.850199&0.635465&18.849929&0.634991&1.964&${8.60\times 10^{8}}^{+5.9E8}_{-8.2E8}$&${1.39\times 10^{-1}}^{+4.0E-2}_{-3.5E-2}$&$4.90\times 10^{40}\pm 5.1E40$&$5.47\times 10^{38}\pm 1.6E38$\\
\rowcolor{white}
SDSS J014529.26+001036.0&e&26.372410&0.177643&26.371791&0.176987&3.245&${1.14\times 10^{9}}^{+7.8E8}_{-1.1E9}$&${3.61\times 10^{-1}}^{+4.1E-2}_{-9.5E-2}$&$7.77\times 10^{40}\pm 6.4E40$&$1.21\times 10^{39}\pm 3.4E38$\\
\rowcolor{Gray}
SDSS J024117.09-001352.3 *&&40.321250&-0.231196&40.320863&-0.232070&3.444&${1.70\times 10^{9}}^{+3.7E8}_{-2.9E8}$&${2.85\times 10^{-1}}^{+1.4E-1}_{-1.1E-1}$&$4.71\times 10^{39}\pm 7.7E39$&$1.10\times 10^{39}\pm 3.6E38$\\
\rowcolor{white}
2MASX J02564580+0603173 &&44.189983&6.055122&44.190282&6.054252&3.308&${3.09\times 10^{6}}^{+3.5E6}_{-1.1E5}$&${3.21\times 10^{-2}}^{+5.2E-2}_{-7.6E-2}$&$3.41\times 10^{39}\pm 1.9E40$&$7.62\times 10^{37}\pm 2.0E37$\\
\rowcolor{Gray}
LEDA 2402319 *&&123.634210&51.884872&123.635037&51.883489&5.309&${2.82\times 10^{9}}^{+5.4E8}_{-4.0E8}$&${6.57\times 10^{-1}}^{+3.2E-1}_{-3.5E-1}$&$3.19\times 10^{40}\pm 4.1E40$&$2.28\times 10^{39}\pm 2.2E39$\\
\rowcolor{white}
2MASX J08193880+2103521 *&i&124.911720&21.064728&124.912299&21.064209&2.694&${1.12\times 10^{9}}^{+2.2E8}_{-2.4E8}$&${5.00\times 10^{-2}}^{+3.5E-1}_{-3.8E-1}$&$2.69\times 10^{39}\pm 4.8E39$&$3.86\times 10^{38}\pm 1.8E38$\\
\rowcolor{Gray}
SDSS J082228.93+034551.7 &&125.620580&3.764374&125.619971&3.764018&2.543&${6.63\times 10^{8}}^{+1.3E8}_{-1.2E8}$&${1.30\times 10^{0}}^{+2.8E-1}_{-2.6E-1}$&$1.13\times 10^{40}\pm 4.9E39$&$3.24\times 10^{39}\pm 3.5E39$\\
\rowcolor{white}
2MASS J08320053+1912058&g&128.002140&19.201637&128.002743&19.201694&2.074&${1.97\times 10^{9}}^{+7.8E8}_{-8.0E8}$&${1.61\times 10^{0}}^{+5.7E-1}_{-4.1E-1}$&$4.16\times 10^{40}\pm 1.8E40$&$4.32\times 10^{39}\pm 7.8E39$\\
\rowcolor{Gray}
SDSS J085629.97+380456.1&&134.124920&38.082253&134.126541&38.080536&7.693&${1.04\times 10^{9}}^{+2.0E8}_{-1.5E8}$&${2.85\times 10^{-1}}^{+3.5E-1}_{-3.1E-1}$&$7.70\times 10^{39}\pm 1.4E40$&$9.42\times 10^{38}\pm 9.5E38$\\
\rowcolor{white}
SDSS J090335.40+151142.0&&135.897540&15.195035&135.897672&15.195871&3.045&${7.49\times 10^{8}}^{+1.7E8}_{-1.5E8}$&${6.70\times 10^{-2}}^{+3.2E-1}_{-3.7E-1}$&$1.46\times 10^{40}\pm 9.8E39$&$3.44\times 10^{38}\pm 2.3E38$\\
\rowcolor{Gray}
SDSSCGB 15.2&&139.676320&16.479506&139.677069&16.479499&2.602&${1.57\times 10^{9}}^{+6.5E8}_{-3.8E8}$&${2.50\times 10^{0}}^{+3.2E-1}_{-2.5E-1}$&$4.24\times 10^{40}\pm 2.9E40$&$6.33\times 10^{39}\pm 7.2E39$\\
\rowcolor{white}
2XMMi J092720.4+362407 *&i&141.835160&36.401897&141.835438&36.402101&1.090&${3.30\times 10^{8}}^{+5.0E7}_{-4.0E7}$&${1.13\times 10^{-1}}^{+2.5E-1}_{-1.9E-1}$&$7.88\times 10^{39}\pm 3.7E39$&$3.44\times 10^{38}\pm 2.4E38$\\
\rowcolor{Gray}
PWC2011 J100805.1+125650&&152.021360&12.947362&152.021634&12.947564&1.199&${2.46\times 10^{9}}^{+5.3E8}_{-4.1E8}$&${3.53\times 10^{-1}}^{+2.2E-1}_{-1.8E-1}$&$1.93\times 10^{40}\pm 1.3E40$&$1.45\times 10^{39}\pm 7.0E38$\\
\rowcolor{white}
SDSS J102526.59+124540.3&&156.360800&12.761314&156.359322&12.760766&5.533&${1.02\times 10^{9}}^{+2.8E8}_{-1.8E8}$&${3.28\times 10^{-1}}^{+3.0E-1}_{-2.1E-1}$&$8.44\times 10^{39}\pm 1.7E40$&$1.03\times 10^{39}\pm 8.4E38$\\
\rowcolor{Gray}
LEDA 30866&&157.254560&29.635242&157.254775&29.635034&1.003&${8.41\times 10^{7}}^{+6.6E7}_{-6.2E7}$&${3.90\times 10^{-1}}^{+5.3E-2}_{-8.9E-2}$&$3.12\times 10^{39}\pm 2.4E40$&$9.49\times 10^{38}\pm 2.8E38$\\
\rowcolor{white}
Mrk 1434 *&c&158.542300&58.063630&158.542303&58.063435&0.698&${1.00\times 10^{7}}^{+4.0E6}_{-4.4E5}$&${6.15\times 10^{-2}}^{+1.3E-1}_{-1.7E-1}$&$1.18\times 10^{40}\pm 3.0E39$&$1.44\times 10^{38}\pm 9.1E37$\\
\rowcolor{Gray}
SDSS J103844.88+533005.2 *&&159.686900&53.501450&159.687208&53.501374&0.703&${3.84\times 10^{7}}^{+1.6E7}_{-1.6E6}$&${2.19\times 10^{-1}}^{+1.9E-1}_{-3.7E-1}$&$3.78\times 10^{39}\pm 2.4E38$&$5.10\times 10^{38}\pm 5.9E38$\\
\rowcolor{white}
LEDA 2116718&&166.425690&38.056496&166.424690&38.056521&2.835&${4.15\times 10^{8}}^{+5.8E7}_{-5.0E7}$&${1.57\times 10^{-1}}^{+2.7E-1}_{-2.3E-1}$&$3.11\times 10^{39}\pm 2.3E40$&$4.73\times 10^{38}\pm 3.9E38$\\
\rowcolor{Gray}
UGC 6192 *&i&167.301650&61.396324&167.301427&61.396035&1.110&${2.43\times 10^{8}}^{+4.1E7}_{-3.0E7}$&${5.04\times 10^{-2}}^{+2.7E-1}_{-2.9E-1}$&$1.60\times 10^{39}\pm 4.2E38$&$1.74\times 10^{38}\pm 1.4E38$\\
\rowcolor{white}
SDSS J112830.77+583342.9 &h&172.128300&58.561844&172.128389&58.561797&0.242&${1.67\times 10^{9}}^{+5.7E8}_{-5.7E8}$&${5.87\times 10^{-1}}^{+4.8E-1}_{-3.9E-1}$&$4.01\times 10^{40}\pm 5.5E39$&$1.76\times 10^{39}\pm 2.5E39$\\
\rowcolor{Gray}
SDSS J112910.56+582309.0 *&i&172.294000&58.385834&172.293456&58.385212&2.468&${1.28\times 10^{9}}^{+2.8E8}_{-2.0E8}$&${1.17\times 10^{0}}^{+3.2E-1}_{-1.4E-1}$&$2.99\times 10^{40}\pm 3.4E40$&$3.12\times 10^{39}\pm 2.7E39$\\
\rowcolor{white}
Mrk 1303 *&i&175.055150&-0.411672&175.055270&-0.411804&0.656&${8.35\times 10^{8}}^{+4.5E8}_{-2.2E8}$&${1.10\times 10^{0}}^{+3.8E-1}_{-1.9E-1}$&$1.06\times 10^{40}\pm 6.7E39$&$2.79\times 10^{39}\pm 3.1E39$\\
\rowcolor{Gray}
2XMM J114501.7+194549 *&&176.257550&19.763748&176.257412&19.763534&0.907&${2.27\times 10^{9}}^{+5.8E8}_{-5.2E8}$&${8.60\times 10^{-2}}^{+4.8E-1}_{-9.4E-1}$&$1.02\times 10^{40}\pm 7.2E39$&$7.65\times 10^{38}\pm 6.2E38$\\
\rowcolor{white}
SDSS J115558.40+232730.7&i&178.993360&23.458563&178.993608&23.459134&2.210&${2.14\times 10^{9}}^{+3.5E8}_{-2.5E8}$&${1.43\times 10^{0}}^{+3.1E-1}_{-2.3E-1}$&$3.80\times 10^{40}\pm 2.1E40$&$4.02\times 10^{39}\pm 3.9E39$\\
\rowcolor{Gray}
NGC 4117 *&i&181.942140&43.126354&181.942229&43.126437&0.380&${2.18\times 10^{9}}^{+4.5E8}_{-4.8E8}$&${1.70\times 10^{-2}}^{+4.3E-1}_{-1.0E0}$&$6.17\times 10^{39}\pm 5.0E38$&$5.53\times 10^{38}\pm 1.5E38$\\
\rowcolor{white}
ECO 11516 *&&182.253720&42.475260&182.253462&42.474715&2.089&${1.92\times 10^{9}}^{+4.4E8}_{-3.8E8}$&${1.10\times 10^{-1}}^{+3.2E-1}_{-3.3E-1}$&$2.21\times 10^{40}\pm 1.3E40$&$7.31\times 10^{38}\pm 3.6E38$\\
\rowcolor{Gray}
SDSS J121352.97+141312.5&&183.470730&14.220132&183.469156&14.220695&5.867&${1.11\times 10^{9}}^{+2.6E8}_{-2.5E8}$&${3.28\times 10^{-1}}^{+7.9E-2}_{-6.5E-2}$&$3.36\times 10^{39}\pm 8.5E39$&$1.05\times 10^{39}\pm 2.4E38$\\
\rowcolor{white}
SDSS J121707.89+034056.3 *&&184.282880&3.682264&184.283938&3.680705&6.775&${1.05\times 10^{7}}^{+5.8E6}_{-9.1E6}$&${5.30\times 10^{-2}}^{+1.5E-1}_{-1.3E-1}$&$2.24\times 10^{39}\pm 1.4E39$&$1.25\times 10^{38}\pm 7.2E37$\\
\rowcolor{Gray}
LEDA 39539&&184.632190&5.849806&184.632586&5.851398&5.909&${2.17\times 10^{8}}^{+4.2E7}_{-5.1E7}$&${1.90\times 10^{-3}}^{+5.4E-1}_{-1.0E0}$&$2.47\times 10^{39}\pm 1.3E39$&$5.62\times 10^{37}\pm 1.7E37$\\
\rowcolor{white}
NGC 4395 *&b; c; f; h; i&186.453610&33.546870&186.453591&33.546854&0.091&${2.50\times 10^{7}}^{+6.4E6}_{-5.6E6}$&${1.57\times 10^{-4}}^{+6.5E-1}_{-7.5E-1}$&$1.39\times 10^{40}\pm 7.0E38$&$6.23\times 10^{36}\pm 1.6E36$\\
\rowcolor{Gray}
2XMM J123519.9+393110&i&188.833530&39.519196&188.832976&39.519672&2.298&${7.85\times 10^{7}}^{+1.5E7}_{-1.2E7}$&${2.60\times 10^{-1}}^{+1.7E-1}_{-1.6E-1}$&$1.10\times 10^{40}\pm 4.4E39$&$6.26\times 10^{38}\pm 4.3E38$\\
\rowcolor{white}
NVSS J123542-001252&&188.927000&-0.215192&188.926617&-0.214698&2.255&${6.47\times 10^{7}}^{+3.2E7}_{-2.6E7}$&${4.15\times 10^{-1}}^{+7.8E-2}_{-5.3E-2}$&$7.14\times 10^{39}\pm 9.3E39$&$9.89\times 10^{38}\pm 2.7E38$\\
\rowcolor{Gray}
LEDA 44693 *&&195.004040&27.945436&195.004703&27.945354&2.118&${1.73\times 10^{9}}^{+3.6E8}_{-4.1E8}$&${2.22\times 10^{-2}}^{+4.6E-1}_{-1.0E0}$&$8.02\times 10^{39}\pm 6.0E39$&$4.77\times 10^{38}\pm 1.8E38$\\
\rowcolor{white}
7W 1258+27W06 *&i&195.140300&27.637766&195.140594&27.637073&2.659&${2.64\times 10^{9}}^{+6.1E8}_{-3.9E8}$&${1.04\times 10^{0}}^{+2.7E-1}_{-1.8E-1}$&$2.45\times 10^{40}\pm 2.3E40$&$3.10\times 10^{39}\pm 2.3E39$\\
\rowcolor{Gray}
2MASX J13070847+5357446 *&i&196.785110&53.962387&196.784717&53.962394&0.835&${2.37\times 10^{9}}^{+9.2E8}_{-8.1E8}$&${2.20\times 10^{0}}^{+1.2E-1}_{-7.4E-2}$&$4.89\times 10^{40}\pm 3.3E40$&$5.78\times 10^{39}\pm 2.1E39$\\
\rowcolor{white}
SDSS J130821.42+113055.0&&197.089300&11.515293&197.087516&11.515225&6.279&${2.03\times 10^{8}}^{+2.5E7}_{-2.9E7}$&${2.17\times 10^{-1}}^{+2.7E-1}_{-1.8E-1}$&$4.99\times 10^{39}\pm 1.5E40$&$5.60\times 10^{38}\pm 4.8E38$\\
\rowcolor{Gray}
SDSS J131930.27+552146.0&&199.876170&55.362810&199.877721&55.362280&3.691&${1.09\times 10^{9}}^{+2.0E8}_{-1.4E8}$&${2.72\times 10^{-1}}^{+2.8E-1}_{-2.6E-1}$&$7.34\times 10^{39}\pm 6.0E39$&$9.04\times 10^{38}\pm 7.2E38$\\
\rowcolor{white}
2XMM J134107.9+263047&a&205.283140&26.513401&205.282967&26.513346&0.596&${2.50\times 10^{9}}^{+1.1E9}_{-1.0E9}$&${1.22\times 10^{0}}^{+2.0E-1}_{-1.0E-1}$&$5.54\times 10^{40}\pm 1.2E41$&$3.67\times 10^{39}\pm 1.9E39$\\
\rowcolor{Gray}
2XMM J134427.6+560130&d&206.114010&56.024930&206.114470&56.025227&1.410&${1.40\times 10^{9}}^{+7.9E8}_{-5.3E8}$&${1.60\times 10^{1}}^{+1.0E-1}_{-8.1E-2}$&$1.57\times 10^{41}\pm 8.2E40$&$3.99\times 10^{40}\pm 1.5E40$\\
\rowcolor{white}
2XMM J134719.1+581437 *&&206.830260&58.243744&206.829865&58.243869&0.876&${1.26\times 10^{9}}^{+2.6E8}_{-1.7E8}$&${5.08\times 10^{-1}}^{+3.2E-1}_{-2.5E-1}$&$1.23\times 10^{40}\pm 5.5E39$&$1.52\times 10^{39}\pm 1.4E39$\\
\rowcolor{Gray}
2XMM J134736.4+173404&d&206.901690&17.567960&206.901721&17.567890&0.278&${2.28\times 10^{9}}^{+6.4E8}_{-3.9E8}$&${1.26\times 10^{0}}^{+4.7E-1}_{-2.8E-1}$&$8.88\times 10^{41}\pm 5.9E40$&$3.59\times 10^{39}\pm 4.7E39$\\
\rowcolor{white}
UGC 9215&&215.862980&1.726289&215.863058&1.726305&0.299&${6.87\times 10^{8}}^{+1.4E8}_{-9.3E7}$&${2.17\times 10^{-1}}^{+3.1E-1}_{-2.6E-1}$&$3.85\times 10^{39}\pm 1.6E39$&$6.61\times 10^{38}\pm 5.9E38$\\
\rowcolor{Gray}
SDSS J143102.57+281625.9&&217.760760&28.273851&217.760817&28.273226&2.258&${9.17\times 10^{8}}^{+2.5E8}_{-1.7E8}$&${3.14\times 10^{-1}}^{+1.7E-1}_{-1.0E-1}$&$8.26\times 10^{39}\pm 9.3E39$&$9.71\times 10^{38}\pm 4.2E38$\\
\rowcolor{white}
2MASX J14401271+0247441 *&b; c; d; f; h; i&220.052920&2.795424&220.052756&2.795508&0.657&${2.66\times 10^{9}}^{+5.6E8}_{-3.9E8}$&${7.20\times 10^{-1}}^{+2.9E-1}_{-2.5E-1}$&$2.42\times 10^{40}\pm 8.1E39$&$2.36\times 10^{39}\pm 1.9E39$\\
\rowcolor{Gray}
2XMM J144056.3+033145&&220.236020&3.528175&220.235329&3.527134&4.501&${1.44\times 10^{8}}^{+6.9E7}_{-8.6E7}$&${1.42\times 10^{-1}}^{+9.7E-2}_{-5.1E-2}$&$4.34\times 10^{39}\pm 8.9E39$&$3.74\times 10^{38}\pm 1.1E38$\\
\rowcolor{white}
ECO 2050 *&&228.550740&13.809550&228.549996&13.810569&4.487&${4.27\times 10^{8}}^{+7.8E7}_{-5.6E7}$&${1.33\times 10^{-1}}^{+2.9E-1}_{-2.8E-1}$&$2.07\times 10^{39}\pm 4.8E39$&$4.17\times 10^{38}\pm 3.8E38$\\
\rowcolor{Gray}
SDSS J153704.18+551550.5 *&&234.267400&55.264060&234.267702&55.263194&3.186&${7.24\times 10^{6}}^{+1.3E6}_{-1.1E6}$&${2.46\times 10^{-2}}^{+2.2E-1}_{-2.3E-1}$&$1.13\times 10^{39}\pm 1.8E38$&$5.82\times 10^{37}\pm 5.4E37$\\
\rowcolor{white}
SDSS J154818.94+350741.2&&237.078920&35.128174&237.079120&35.128482&1.259&${1.66\times 10^{9}}^{+3.3E8}_{-2.2E8}$&${5.82\times 10^{-1}}^{+2.7E-1}_{-2.0E-1}$&$7.10\times 10^{39}\pm 1.4E40$&$1.85\times 10^{39}\pm 1.4E39$\\
\rowcolor{Gray}
2XMM J160531.8+174825 *&b; f; i&241.382700&17.807276&241.382749&17.807222&0.245&${1.64\times 10^{9}}^{+3.5E8}_{-3.5E8}$&${1.53\times 10^{-1}}^{+2.5E-1}_{-2.3E-1}$&$6.91\times 10^{40}\pm 3.2E40$&$7.73\times 10^{38}\pm 3.8E38$\\
\rowcolor{white}
AGC 262533 *&i&241.712550&8.157967&241.712419&8.158277&1.218&${1.64\times 10^{8}}^{+2.5E7}_{-2.0E7}$&${4.94\times 10^{-2}}^{+1.8E-1}_{-1.4E-1}$&$3.16\times 10^{39}\pm 9.4E38$&$1.53\times 10^{38}\pm 7.7E37$\\
\rowcolor{Gray}
SDSS J161321.26+510534.8&&243.338610&51.093000&243.340961&51.094039&6.507&${1.57\times 10^{9}}^{+2.2E8}_{-1.8E8}$&${6.25\times 10^{-1}}^{+2.8E-1}_{-2.4E-1}$&$1.09\times 10^{40}\pm 2.0E40$&$1.88\times 10^{39}\pm 1.6E39$\\
\rowcolor{white}
SDSS J162642.49+390842.8 *&&246.677060&39.145226&246.676418&39.145578&2.203&${7.65\times 10^{8}}^{+1.7E8}_{-1.0E8}$&${3.90\times 10^{-1}}^{+2.3E-1}_{-1.3E-1}$&$3.04\times 10^{40}\pm 1.4E40$&$1.11\times 10^{39}\pm 6.9E38$\\
\rowcolor{Gray}
SDSS J162729.77+385455.1&&246.874050&38.915330&246.873416&38.915903&2.733&${1.08\times 10^{9}}^{+2.4E8}_{-1.7E8}$&${2.53\times 10^{-1}}^{+1.8E-1}_{-1.2E-1}$&$1.60\times 10^{40}\pm 1.8E40$&$8.69\times 10^{38}\pm 3.8E38$\\
\rowcolor{white}
SDSS J213732.54+002800.1&&324.385620&0.466721&324.386684&0.466883&3.874&${6.87\times 10^{8}}^{+2.0E8}_{-1.5E8}$&${2.56\times 10^{-1}}^{+3.6E-1}_{-3.8E-1}$&$3.95\times 10^{39}\pm 8.9E39$&$7.97\times 10^{38}\pm 9.6E38$\\
\rowcolor{Gray}
SDSS J213743.69+003125.5&&324.432130&0.523779&324.432454&0.523861&1.206&${1.19\times 10^{9}}^{+2.7E8}_{-2.1E8}$&${1.87\times 10^{-1}}^{+2.6E-1}_{-2.2E-1}$&$2.34\times 10^{40}\pm 2.2E40$&$7.63\times 10^{38}\pm 4.7E38$\\
\rowcolor{white}
6dFGS gJ233225.3-005049 *&&353.105260&-0.847034&353.106172&-0.848164&5.240&${8.26\times 10^{8}}^{+1.1E8}_{-8.4E7}$&${3.65\times 10^{-1}}^{+2.6E-1}_{-2.6E-1}$&$1.35\times 10^{40}\pm 9.8E39$&$1.05\times 10^{39}\pm 9.2E38$\\
\rowcolor{Gray}
SDSS J011421.73+001335.6&&18.590588&0.226537&18.590958&0.228235&6.258&${1.92\times 10^{9}}^{+3.1E9}_{-4.0E9}$&${2.71\times 10^{-2}}^{+3.6E-3}_{-3.6E-3}$&$1.59\times 10^{41}\pm 6.2E41$&$6.63\times 10^{38}\pm 9.1E38$\\
\rowcolor{white}
SDSS J012325.32-002921.4 *&&20.855507&-0.489282&20.854883&-0.488693&3.088&${8.53\times 10^{8}}^{+6.3E8}_{-9.3E8}$&${1.14\times 10^{-3}}^{+9.4E-5}_{-9.4E-5}$&$5.02\times 10^{39}\pm 1.3E40$&$2.15\times 10^{38}\pm 1.6E38$\\
\rowcolor{Gray}
SDSS J030446.14-011208.1 *&&46.192287&-1.202276&46.192499&-1.204069&6.501&${1.52\times 10^{8}}^{+1.1E8}_{-1.6E8}$&${4.85\times 10^{-4}}^{+3.8E-5}_{-3.8E-5}$&$5.66\times 10^{40}\pm 2.6E40$&$3.81\times 10^{37}\pm 2.8E37$\\
\rowcolor{white}
SDSS J220558.60-003049.3&&331.494140&-0.513736&331.495163&-0.513464&3.809&${2.92\times 10^{9}}^{+2.3E9}_{-3.4E9}$&${2.11\times 10^{-2}}^{+1.2E-3}_{-1.2E-3}$&$1.27\times 10^{41}\pm 8.6E40$&$8.64\times 10^{38}\pm 6.7E38$\\
\rowcolor{Gray}
SDSS J234759.26+010344.2&&356.996920&1.062305&356.996524&1.061061&4.697&${1.27\times 10^{9}}^{+2.5E9}_{-2.9E9}$&${7.23\times 10^{-2}}^{+2.1E-2}_{-2.1E-2}$&$5.35\times 10^{42}\pm 5.2E42$&$7.09\times 10^{38}\pm 8.4E38$

\end{longtable}
\twocolumn
\end{landscape}
\end{small}

\bsp	\label{lastpage}
\end{document}